\documentclass[fleqn,usenatbib]{mnras}

\usepackage{newtxtext,newtxmath}

\usepackage[T1]{fontenc}

\DeclareRobustCommand{\VAN}[3]{#2}
\let\VANthebibliography\thebibliography
\def\thebibliography{\DeclareRobustCommand{\VAN}[3]{##3}\VANthebibliography}

\usepackage{graphicx}	
\usepackage{amsmath}	
\usepackage[table]{xcolor} %
\usepackage{amsmath}
\usepackage{graphicx}
\usepackage{booktabs}
\usepackage{geometry}
\title[ Structure growth in  $f(Q)$ cosmology]{ Structure growth in  $f(Q)$ cosmology}

\author[S. Sahlu et al.]{
Shambel Sahlu,$^{1,2}$\thanks{sahlushambel@gmail.com}
\'{A}lvaro de la Cruz-Dombriz,$^{3,4}$ \thanks{alvaro.dombriz@gmail.com} and 
Amare Abebe$^{1,5}$ \thanks{amareabebe.gidelew@nwu.ac.za}
\\
$^{1}$Centre for Space Research, North-West University, Potchefstroom, South Africa\\
$^{2}$Department of Physics, Wolkite University, Wolkite, Ethiopia\\
$^{3}$Cosmology and Gravity Group, Department of Mathematics and Applied Mathematics, University of Cape Town, 7700 Rondebosch, South Africa\\
$^4${Departamento de F\'isica Fundamental, Universidad de Salamanca, 37008 Salamanca, Spain}\\
$^5${National Institute for Theoretical and Computational Sciences (NITheCS), South Africa}
}

\date{Accepted XXX. Received YYY; in original form ZZZ}

\pubyear{\the\year{}}

\begin{document}
\label{firstpage}
\pagerange{\pageref{firstpage}0.00\pageref{lastpage}}
\maketitle
\renewcommand{\arraystretch}{1.3} 
\setlength{\tabcolsep}{6pt} 
\begin{abstract}
We take into account redshift space distortion measurements to investigate the growth of cosmological large-scale structures within the framework of {generalised} symmetric teleparallel - $f(Q)$ - gravity. 
After comparing the predictions of the $f(Q)$-gravity expansion history with OHD and {SNIa Pantheon+ sample} datasets and constraining the pertinent cosmological parameters $\Omega_{m}$ and $H_0$, together with the exponent $n$ for $f(Q)$ power-law models, we derive the full system of equations governing linear cosmological perturbations to study matter fluctuations using the $1 + 3$ covariant formalism when applied to $f(Q)$ gravity. 
Thus, we resort to both the usual redshift-space distortion data $f\sigma_8$ and some recent separate measurements of the growth rate $f$ and the amplitude of matter fluctuations $\sigma_8$ from the VIPERS and SDSS collaborations to find the best-fit cosmological parameters $\Omega_m$, $\sigma_{8}$ and $n$. We also apply a collective analysis of such growth-structure data together with the aforementioned cosmic expansion measurements, to {constrain} these parameters through Monte Carlo Markov Chain simulations.
As a result, our analysis is capable of determining the statistical significance for the best-fit parameter values through the \rm{AIC} and \rm{BIC} Bayesian selection criteria. 
\end{abstract}


\section{Introduction}\label{sec:intro}
Extensions of general relativity (GR) have drawn significant interest in explaining the late-time accelerated expansion history of the universe \citep{riess1998observational} with no requirement for a scenario that includes dark energy {\it ad hoc}. Such attempts have also drawn great attention in the realm of relativistic cosmology to solve many unsolved problems such as inhomogeneities and anisotropies, matter density fluctuations, and the formation of large-scale structures, to mention just a few.  The recent observational data presented in \citep{ reichardt2012measurement} showed that our observable universe is experiencing an accelerating expansion, perhaps the clearest clue to the properties of gravity beyond GR. The main pillar of the mystery of cosmic expansion
is that a major part of the universe is made up of a substance known as dark energy \citep{barris2004twenty} or extended - also dubbed modified - gravity on cosmological scales. The Einstein-Hilbert action is described by $f(Q)$ - $Q$ being the nonmetricity scalar \citep{Capzilo}, $f(T)$ - $T$ being the torsion scalar \citep{Cai}, $f(R)$ - $R$ being the Ricci scalar \citep{De}, a combination of the two dubbed $f(R, T)$ \citep{Myrzakulov} or $f(G)$ - $G$ being the Gauss-Bonnet invariant \citep{De1}, among others.  
\\
\\
 One of the recent additions to the myriad of modifications to GR, $f(Q)$-gravity theory is based on the generalisation of the symmetric teleparallel equivalent of GR (STEGR), characterised by a connection in which both the curvature and torsion vanish. Consequently, STEGR is described by a non-metric scalar $Q$, representing one of the equivalent geometric formulations of GR \citep{ Khyllep}. Thus, $f(Q)$ has gained much attention to study different aspects of cosmology
and parameter constraints \citep{DAgostino}. The detailed clarifications of STEGR when referring to the background evolution of the universe have been studied in \citep{ayuso2021observational} whereas the different self-accelerating solutions have been studied in \citep{Junior}, all of resorting to the predictions of the cosmological expansion history within this class of theories resorting solely to 
observations such as high-redshift Hubble diagrams from SNIa, baryon acoustic oscillations (BAO) or CMB shift factor, all of them based on different distance measurements which are only sensitive to the expansion history.
\\
\\
Nevertheless, there exist a different type of observations that are sensitive not only to the expansion history but also to the evolution of matter density perturbations. 
The fact that the evolution of perturbations depends on the specific gravity theory, i.e., it differs in general from that of Einstein’s gravity even though the background evolution is the same, has made this kind of
observations instrumental in distinguishing between different cosmological models and the mechanisms that cause cosmological acceleration.
Within this context, there exist two basic approaches to cosmological perturbations in the literature: the first is the gauge-invariant perturbations, also dubbed metric, formalism \citep{kodama1984cosmological}, whereas the second approach is the $1+3$ covariant gauge-invariant perturbations formalism \citep{dunsby1992cosmological} which will be employed in this work.  
The first approach presents several shortcomings such that it does not provide a covariant description \citep{w}, and can be nonlinear because it is built around a local comoving observer frame. 
The second approach though is covariant and gauge invariant. Also, as widely known, it uses the kinematic quantities, the energy-momentum tensor of the fluid(s), and the gravito-electromagnetic parts of the Weyl tensor.
Also, the second approach to perturbations theory produces differential equations that help to describe the true physical degrees of freedom. As an example, different works have used the $1+3$ covariant formalism in GR \citep{dunsby1991gauge}, $f(R)$ \citep{abebe2012covariant,ntahompagaze2018study} and, in the context of $f(T)$ 
\citep{sami2021covariant, sahlu2020scalar} 
to investigate the evolution of large-scale structures and to compute the gravitational instabilities seeding the formation of stars, quasars, galaxies, and clusters \citep{fry1984galaxy}. 
\\
\\
Amongst those cosmological observations that are sensitive to the evolution of matter density perturbations, the fact that the spatial distribution of galaxies appears squashed and distorted \citep{Percival} when their positions are shown as functions of their redshift rather than as functions of their distance, plays an instrumental role in observational cosmology nowadays. This is the well-known \textit{redshift-space distortion} effect.  It is possible to measure redshift space distortions to set constraints on cosmological parameters, and as a byproduct, it is also possible to integrate these distortions to recoup the underlying actual space correlation function. In this regard, data on growth rate measurements $f\sigma_8 = \sigma_8\frac{\rm d\ln{\mathcal{D}}}{\rm d\ln{a}}$ have recently been disclosed, where $a$ is the cosmological scale factor, $\mathcal{D}$ is the linear growth of matter overdensity and $\sigma_8$ is the normalisation of the matter power spectrum on the scale of $8h^{-1}{\rm Mpc}$. As a consequence, such growth-structure data have been considered to study different modified theories of gravity. For instance, just to cite a few efforts, in \citep{Albuquerque} the growth of structures in $f(Q)$ gravity has been performed under the metric formalism by following the designer approach.  In \citep{Jennings} $N$-body simulations in $f(R)$ gravity enabled the authors to make redshift-space predictions on the dark-matter clustering such theories. Finally, the use of growth structure data in the context of the effective field theory of dark energy formalism presented allowed authors in \citep{Perenon} to constrain the cosmological parameters. 
\\
 \\
 In this work, we shall first pay attention to the cosmological expansion history resorting to both OHD, SNIa and combined OHD+SNIa data to constrain the cosmological parameters for power-law gravity  $f(Q) = Q+F(Q)$ models \citep{Khyllep}, where $F(Q) = \alpha Q^n$. {For the case of $n=0$, this gravity model reduces to $\Lambda$CDM.}
%
In this communication, both the MCMC hammer \citep{foreman2013emcee} and GetDist \citep{lewis2019getdist} software packages shall be used to constrain these best-fit parameters \{$\Omega_m$, $H_0$, $n$\} for $f(Q)$ gravity model by using { SNIa distance moduli measurements from the Pantheon+ sample \cite{brout2022pantheon+}, which consists of 1701 light curves of 1550 distinct SNIa ranging in the redshift interval $z \in [0.001, 2.26]$.
%
%
We shall also use the observational Hubble parameter data (OHD), from Baryon acoustic oscillations (BAO) provided by the Dark Energy Spectroscopic Instrument (DESI) 2024 Survey \cite{adame2024desi} and cosmic chronometers or differential ages of galaxies at the redshift range $0.0708 < z \leq 2.36$, \citep{yu2018hubble}.}
\\
\\
Second, after studying the background cosmology in the framework of $f(Q)$ gravity, the crucial point in this investigation is to perform a detailed analysis of linear cosmological perturbations using the $1+3$ covariant formalism to study the growth of large-scale structures. Therefore, we shall consider both the whole system and the so-called quasistatic approximation of the perturbation equations. This will allow us to better understand the limitations of the quasi-static approximation when studying the growth density fluctuations in the realm of $f(Q)$ theories. For further tasks, 
we have implemented the sets of redshift-space distortion data $f\sigma{_8}$ with the three latest separate measurements of the growth rate $f$  and the amplitude of matter fluctuations $\sigma_8$ from the VIMOS Public Extragalactic Redshift Survey (VIPERS)  and SDSS collaborations.
In particular, we will compare constraints obtained from three datasets:  
$i)$ the combinations of the measurements of  the growth rate ($f$), and $\sigma_8$, sourced from SDSS and Vipers PDR-2 Collaborations \citep{De11}, collectively labeled as $\sigma_8+f$; 
$ii)$  the  30  redshift-distortion measurements of $f\sigma_8$, dubbed $f\sigma_8$, covering the redshift range $0.001\leq z\leq 1.944$ and; 
$iii)$ the full combinations of thirty $f\sigma_8$, three combined three $f$, and three $\sigma_8$ measurements collectively labeled as $\sigma_8+f+f\sigma_8$ (for further details refer to Table \ref{datasetsofgrowth1} and \ref{datasetsofgrowth} in Appendix \ref{Appendix}). For the latter case $iii)$, the $f\sigma+8$ measurements with their $f$ and $\sigma_8$ counterparts are indeed not taken into account for the independence of the data. { Using MCMC simulations, these sets of data - either by themselves or combined with OHD and SNIa measurements - will allow us to constrain the corresponding best-fit values of $\{ H_0,  \Omega_m, \sigma_{8}, n\}$. In order to do so, throughout this manuscript, the following priors have been assumed: \( H_0 = [50.00, 80.00]\) in \({\rm km\, s^{-1}\,Mpc^{-1}},\, \Omega_m = [0.00, 1.00]\), \(\sigma_8 = [0.50,1.00] \) and $n = [-0.50, 0.50]$, for the power-law $f(Q) = Q+\alpha Q^n$  gravity model under study in what follows.}
\\
\\
{At this stage, a pertinent comment refers to the crucial difficulty 
teleparallel gravity theories, namely \( f(G) \) and \( f(T) \) and $f(Q)$, face in balancing strong couplings due to ghost instabilities. For instance, in the context of \( f(Q) \) theories, the choice of connection affects the number of degrees of freedom and the overall stability. Such a choice can help reduce the likelihood of ghost formation. As discussed in \citet{aguiar2024pathological}, two of those connections\footnote{In this communication we have chosen one of those two, \(C_1\) following the terminology in 
\citet{aguiar2024pathological}.} exhibit suppressed linear spectra, indicating they are strongly coupled. For the remaining connection, such an analysis disclosed the presence of seven gravitational degrees of freedom, with at least one identified as a ghost. 
A potential solution to these problems could involve exploring a direct coupling between the connection and matter fields, or a non-minimal coupling, as suggested by \citet{heisenberg2024cosmological} and exemplified in \citet{Bello-Morales:2024vqk}. In our study, we have decided nonetheless to consider a gravitational action with minimal coupling between the matter and the connection, as described in Eq. \eqref{eq1} below, even if ghost issues may emerge. Our goal is thus twofold: first to provide a full derivation of the $1+3$ formalism applicable to scalar perturbations for $Q$-based theories, which can be easily extended for different couplings; and second to 
conclude, that even without invoking the ghost's presence, 
the competitiveness of power-law $f(Q)$ models against $\Lambda$CDM remains limited. 
}
 \\
 \\
The layout of the manuscript is as follows: in Section \ref{gravity}, the cosmology of $f(Q)$ gravity is reviewed; the cosmological parameters are constrained, and the viability of the $f(Q)$ gravity model is statistically analysed using OHD, SNIa, and OHD+SNIa datasets. Also, some caveats of these models are presented.
Then,  in Section \ref{gravity1} the $f(Q)$-gravity first- and second-order linear evolution equations using the $1+3$ covariant formalism are derived. This section applies scalar and harmonic decomposition techniques to find solutions for density fluctuations. We also present the theoretical results of the matter density fluctuations when applied to the power-law $f(Q)$-gravity models. 
Subsequently, Section \ref{sigma} is devoted to comparing the linear growth of the matter density predictions against $\sigma_8+f$, $f\sigma_8$ and $\sigma_8+f+f\sigma_8$ and presents the statistical significance for $f(Q)$-gravity power-law models. Also for these models, an exhaustive joint analysis of theoretical predictions against growth structure data and cosmic expansion measurements, i.e., OHD and SNIa data, is performed here.
Finally, we conclude with the conclusions of the manuscript in Section \ref{gravity2}. 

\section{Background cosmology of $f(Q)$-gravity}
\label{gravity}
{
The action of non-metric gravity as presented in \citep{Khyllep,capozziello2022comparing}
\begin{equation}\label{eq1}
	S=\int\sqrt{-g} \left(\frac{1}{2}f(Q)+\mathcal{L}_m\right) {\rm d}^4x, 
\end{equation}
where $f(Q)$ is an arbitrary function of the nonmetricity $Q$, ${g}$ is the determinant of the metric $g_{\mu\nu}$ and $\mathcal{L}_m$ is the matter Lagrangian density\footnote{Unless otherwise indicated, throughout the manuscript we use $c=8\pi G =1$ units.}. 
Variations of action \eqref{eq1} with respect to the metric tensor yields the field equations as
\begin{eqnarray}
    && \frac{{2}}{\sqrt{-g}}\nabla_\gamma \left( \sqrt{-g}f_{Q} P^\gamma_{\mu\nu}\right)+\frac{1}{2}g_{\mu\nu}f \nonumber\\&& +f_Q\left( P_{\mu\nu\delta} Q_v ^{\gamma\delta} -2Q_{\gamma\delta\mu}P^{\gamma \delta}_{\nu}\right) = -T_{\mu\nu}\;,\nonumber\\
\end{eqnarray}
where   $T_{\mu\nu} = -\frac{2}{\sqrt{-g}}\frac{\delta(\sqrt{-g}\mathcal{L}_m)}{\delta g^{\mu\nu}}$ is the energy-momentum tensor and the nonmetricity tensor is the covariant derivative of the metric  tensor
\begin{eqnarray}
Q_{\gamma\mu\nu} \equiv \nabla_\gamma g_{\mu\nu}\;,
\end{eqnarray}
whose traces are  
\begin{eqnarray}
Q_\gamma = Q_{\gamma}~^{\mu}~_{\mu} \qquad \tilde{Q}_\gamma = Q^{\mu}~_{\gamma\mu}\,,
\end{eqnarray}
and the superpotential term 
$P^\gamma_{\mu\nu}$ is  
\begin{eqnarray}
&& P^\gamma_{\mu\nu} = \frac{1}{4}\Bigg( - Q^\gamma~_{\mu\nu}+2Q_{(\mu}~^\gamma~_{\nu )} - Q^\gamma g_{\mu\nu} \nonumber\\&& - \tilde{Q}^\gamma g_{\mu\nu} - \delta ^\gamma_{(\mu}Q_{\nu)}\Bigg)\;,
\end{eqnarray}
with
\begin{eqnarray}
\qquad Q = -Q_{\mu\nu\gamma}P^{\mu\nu\gamma}\;.
\end{eqnarray}
}
In the following, we shall assume a spatially flat Friedmann-Lema\^{i}tre-Robertson-Walker (FLRW) metric
  \begin{equation}
      {\rm d}s^2 = -{\rm d}t^2 +a^2(t)\delta_{\alpha\beta}{\rm d}x^\alpha {\rm d}x^\beta\;,
  \end{equation}
  where $a(t)$ is the cosmological scale factor and $\delta_{\alpha\beta}$ the Kronecker delta. Thus, the trace of the nonmetric tensor becomes $Q = 6H^2$ and the corresponding modified Friedmann and Raychaudhuri equations, once the gravitational Lagrangian has been split as $f(Q) = Q +F(Q)$ become
\begin{eqnarray}
	3H^2= \rho_m + \rho_r+\frac{F}{2}-QF'\label{33ab}\;,
\end{eqnarray}
\begin{eqnarray}\label{333}
	2\dot{H}+3H^2=-\frac{\rho_{r}}{3}-\frac{F}{2}+QF' + \frac{H\dot{Q}}{Q}(2QF''+F')    \label{333}\;,
\end{eqnarray}
respectively{. Here $\rho_m$ and $\rho_r$ are } the energy density of non-relativistic matter and (relativistic matter (radiation) \footnote{Since the contribution of radiation to the late-time cosmological expansion history is so minimal, we have safely neglected such a contribution in the analysis to perform.}, $F'$ and $F''$ are the first and second derivative of $F$ with respect to $Q$ respectively, and the thermodynamic quantities for the effective $f(Q)$ fluid are defined as
  \begin{eqnarray}
	&&\rho_Q=\frac{F}{2}-QF',\label{55}\\&&
	p_Q = -\rho_Q + \frac{\theta \dot{Q}}{3Q}(2QF''+F')\;, \label{555}
\end{eqnarray}
where {$\theta \equiv 3H$}. In the rest of this investigation, we use a generic power-law $F(Q)$-gravity model  
\begin{equation}
\label{model}
	F(Q) = \alpha\, Q^n\;,
\end{equation}
where $n$ is a constant {such that} for $n = 0$, the function $f(Q)$ reduces to $\Lambda$CDM. Similarly, when $n = 1/2$, the above modified Friedmann and Raychaudhuri equations are simplified to {those of} GR. From Eq. \eqref{33ab}, we obtain $\alpha$\footnote{In the following we shall assume $n\neq 1/2$, i.e., $f(Q)$ power-law models strictly different from matter-only GR ($\Omega_m=1$) in order to avoid expression \eqref{modelalpha} being indeterminate.} as
\begin{equation} \label{modelalpha}
\alpha = \frac{1-\Omega_m}{(6H_0^2)^{n-1}(1-2n)}\;.  
\end{equation}
Then with the choice in \eqref{model} and using \eqref{modelalpha}, the normalized Hubble parameter - for $n\neq 1/2$ - yields, 
\begin{eqnarray}
    E^2 \equiv \frac{H^2(z)}{H_0^2} = {\Omega_m}(1+z)^3+(1-\Omega_m)E^{2n}\,,
\end{eqnarray}
where $H_0$ is today's value of the Hubble parameter and  $\Omega_m$ is today's value of the normalized matter (dust) energy density.

%
%
\subsection{Constraining parameters}
As mentioned above, this work will prioritise studying the role of $f(Q)$ gravity at background and perturbation levels compared to the data. In particular, this section focuses on comparing expansion history data with the expansion history predicted by power law $f(Q)$ gravity models in order to elucidate the statistical consistency and soundness of the latter when compared with the $\Lambda$CDM Concordance Model.  

\begin{figure}
    \includegraphics[width=.4\textwidth]{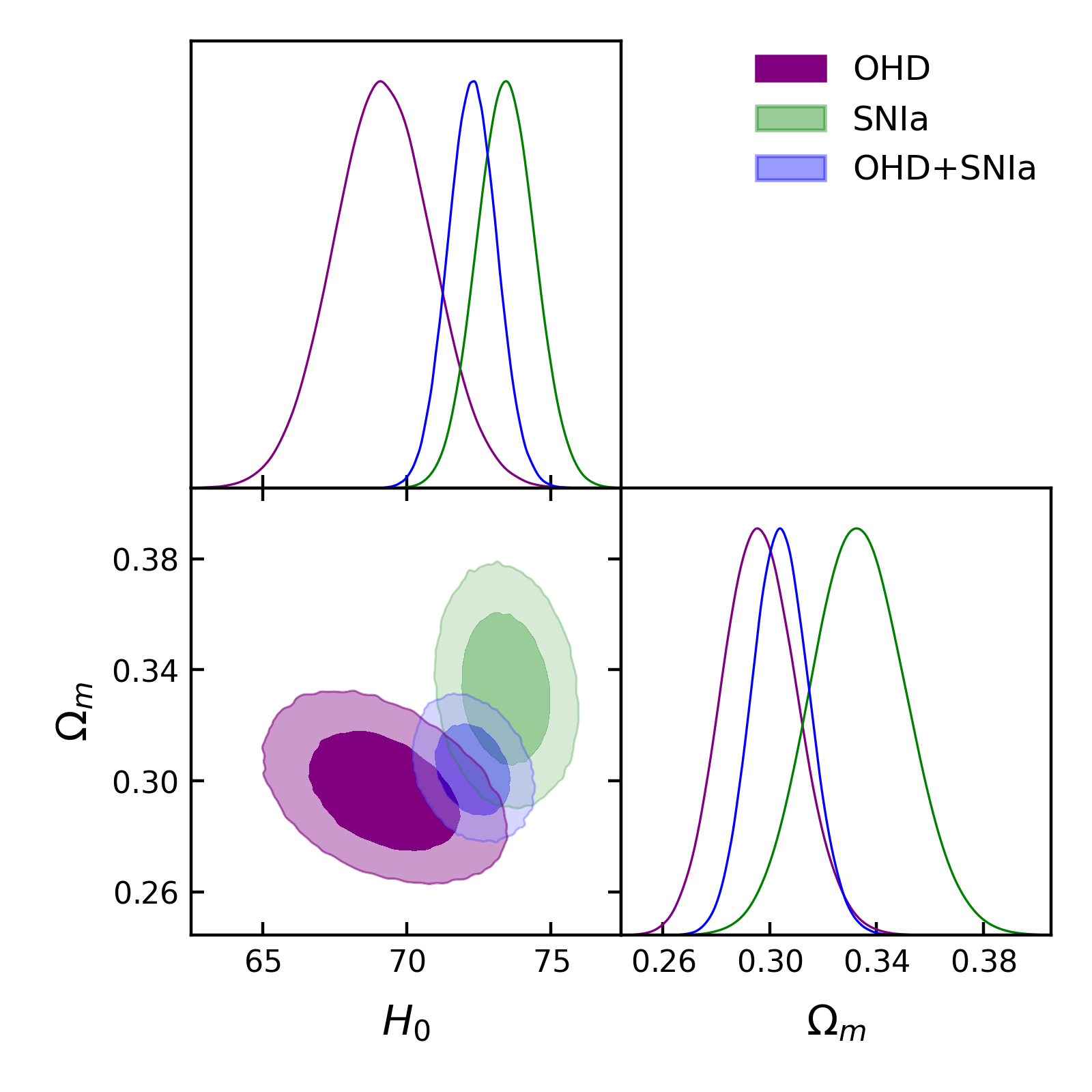}
    \includegraphics[width=.47\textwidth]{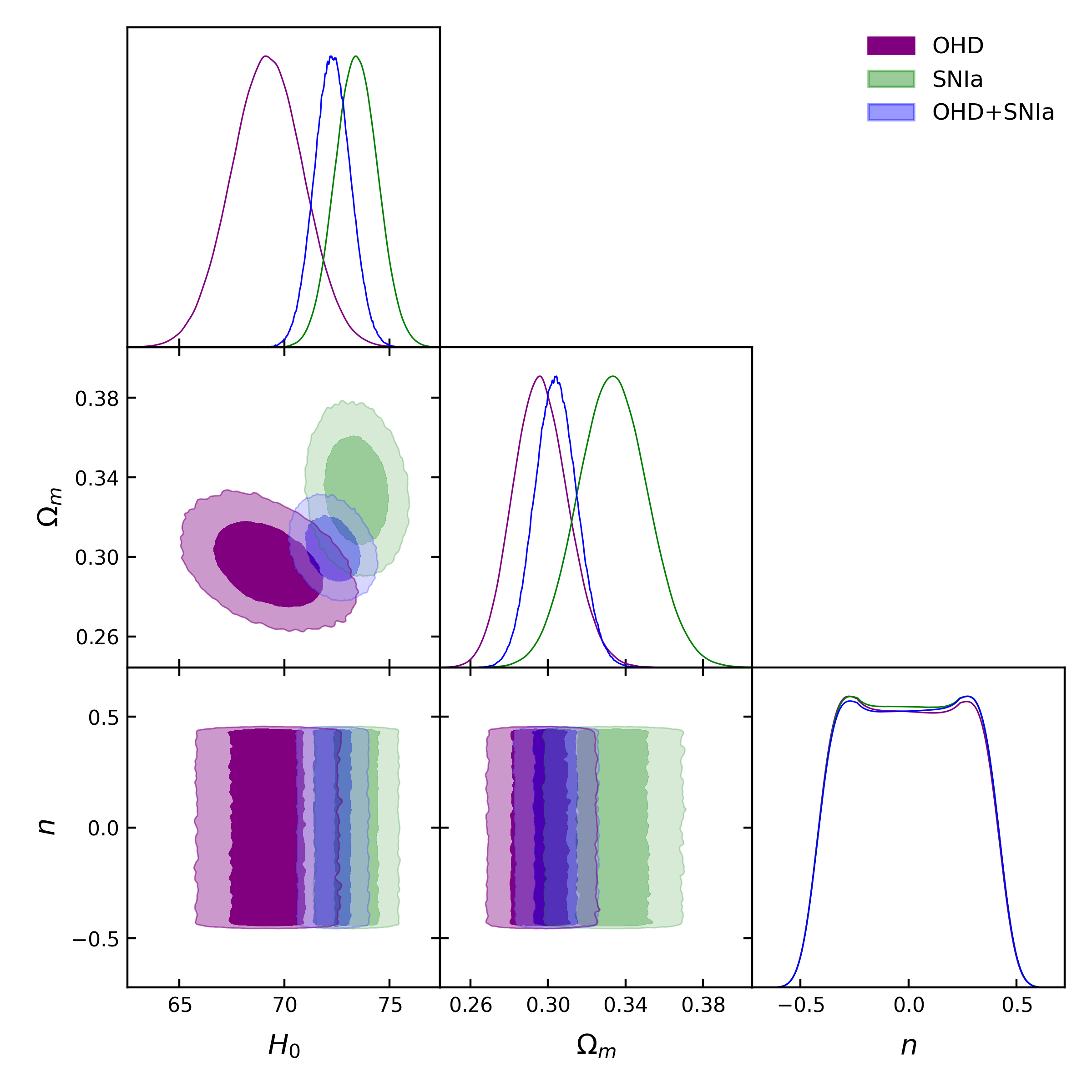}
    \caption{{Best-fit values of the parameters $\Omega_m$ and $H_0$ for the $\Lambda$CDM model (top panel). In the bottom panel, the best-fit values of $\Omega_m$, $H_0$, and $n$ for the $f(Q)$ gravity power law model are shown. The values were calculated using combined datasets OHD, SNIa, and OHD+SNIa at $1\sigma$ and $2\sigma$ confidence levels. For each theory - $\Lambda$CDM or $f(Q)$ - we observe that the OHD and SNIa contour plots are consistent at the $2\sigma$ level but not at the $1\sigma$ level. }
    }
    \label{fig:enter-labelz}
\end{figure}

In doing so, Monte Carlo Markov chain simulations (MCMC) have been implemented taking into account the SNIa, OHD, and OHD+SNIa datasets and contour plots presented in Fig. \ref{fig:enter-labelz} for both $\Lambda$CDM and the $f(Q)$ power-law gravity model for the $1\sigma$ and $2\sigma$ confidence levels.\footnote{Throughout this communication, the MCMC-generated contour plots have used: number of random walkers = $10^3$, number of steps each walker takes = $3\cdot10^4$, and burning the first N-steps = $10^3$ for the Emcee-package initial conditions.}
The best-fit parameters of $H_0$, $\Omega_{m}$, and $n$ are then presented in Table \ref{Tableone}. {The predicted values of  $H_0$ in $f(Q)$ gravity model, specifically using joint datasets, the value of $H_0$ for  $f(Q)$ gravity model is  \(72.85^{+2.220}_{-1.660} ~~\text{km/s/Mpc}\). A detailed comparison of our $H_0$ obtained values with the different both early and late-time measurements will be presented in Section \ref{sigma} Figure \ref{fig:enter-labelH0}.}
    

\begin{table}

    \begin{tabular}{ |c|c|c|c|c|c|c|}
    \hline
Dataset &Parameters & $\Lambda$CDM & $f(Q)$ gravity \\
	\hline
OHD& $n$ & - & {$0.010^{+0.328}_{-0.381}$} \\
&	$\Omega_{m}$ & {$0.287^{+0.039}_{-0.035}$}& {$0.297^{+0.040}_{-0.036}$}\\
&	$H_{0}$\,(${\rm km\,s^{-1} Mpc^{-1}}$) & {$69.72^{+1.92}_{-1.70}$} & {$69. 26^{+3.20}_{-2.70}$} \\
	\hline
 SNIa & $n$ & - & {$-0.080^{+0.324}_{-0.283}$}\\
	&$\Omega_{m}$ & {$0.336^{+0.048}_{-0.045}$} & {$0.332^{+0.054}_{-0.042}$} \\
	& $H_{0}$\,(${\rm km\,s^{-1} Mpc^{-1}}$) & {$73.42^{+1.26}_{-1.52}$} & {$73.26^{+1.60}_{-2.11}$} \\
	\hline
 OHD+SNIa& $n$ & - &{$-0.085_{-0.190}^{+0.300}$}\\
 &$\Omega_{m}$ &{$0.323^{+0.089}_{-0.041}$}&{$0.304^{+0.029}_{-0.027}$}\\
 & $H_0$\,(${\rm km\,s^{-1} Mpc^{-1}}$)&{ $72.56^{+0.60}_{-0.60}$}&{$72.85^{+2.22}_{-1.66}$}\\
 \hline
\end{tabular}
    \caption{{Best-fit values for parameters {$\Omega_m, H_0, n$} {at $1\sigma$ confidence level} when determined for both theories, $\Lambda$CDM and $f(Q)$ power-law, by using three datasets: OHD, SNIa, and OHD+SNIa. } 
    %
    %
    }
    \label{Tableone}
\end{table}
\begin{table*}
		\begin{tabular}{|lllllllll|}
			\hline
			\textbf{Data }&  & \textbf{$\mathcal{L}(\hat{\theta}|{\rm data})$} & \textbf{$\chi ^{2}$} &$\chi ^{2}$-red & \textbf{$\rm{AIC}$} &\textbf{$\Delta \rm{AIC}$} & \textbf{$\rm{BIC}$} & \textbf{$\Delta \rm{BIC}$}\\
   \hline
  $\Lambda{\rm CDM}$& &&&&&&&\\ & &&&&&&&\\
			OHD   & & -14.87 & 29.74 & 0.74 &33.74& -- & 37.26 & --\\
  SNIa && -761.56& 1523.12  &  0.897 & 1527.12 & -- & 1536.99 & --\\
   OHD + SNIa && -781.82& 1563.64 &  0.897 & 1567.64& -- & 1578.57 & --\\
			\hline
    $f(Q)$ model& &&&&&&&\\ & &&&&&&&\\
   OHD &  &-14.60& 29.20& 0.78 & 35.20 & 1.46& 40.48& 3.21\\
   SNIa & &  -761.93& 1522.86 & 0.86 & 1528.86 & 1.74 & 1544.20& 7.20\\
   OHD + 
   SNIa & &  -781.03 & 1562.06 & 0.95 & 1569.06 &1.42 & 1584.45 & 5.88\\
			\hline
	\end{tabular}
	\caption{{Values of $\mathcal{L}(\hat{\theta}|\text{data})$,  $\chi ^{2}$, $\chi^{2}$-red(uced), $\rm{AIC}$, $\Delta\text{AIC}$, $\rm{BIC}$, and $\Delta\text{BIC}$ for both models $\Lambda$CDM and $f(Q)$-gravity power-law models using OHD, SNIa, and OHD+SNIa datasets. For the $f(Q)$-gravity power-law model, values of $\Delta\text{AIC} < 2$ indicate the model has strong observational support; in contrast, values of $\Delta\rm{BIC}$ falling between 3.21 and 7.20 refer to less observational support.}}
 \label{stastical1}
\end{table*} 
\subsection{Statistical analysis} 
We provide statistical analysis using the Akaike information criterion ($\rm{AIC}$) and Bayesian / Schwartz information criterion ($\rm{BIC}$) model selection criterion to compare the viability of $f(Q)$ gravity models compared to $\Lambda$CDM. We take into account the statistical calculations BIC and AIC as presented in \citep{liddle2009statistical,szydlowski2015aic,rezaei2021comparison}  to quantify the extent the power-law $f(Q)$ gravity model should be "accepted" or "rejected" compared to $\Lambda$ CDM. For comparative purposes, we consider the $\Lambda$CDM as the "accepted" model to justify the power-law $f(Q)$ gravity model based on the $\rm{AIC}$ and $\rm{BIC}$ criteria. These criteria allow us to establish the acceptance or rejection of the $f(Q)$-gravity model. The $\rm{AIC}$ and $\rm{BIC}$ values in the $\Lambda$CDM and $f(Q)$ gravity models are calculated considering the following formulation,
\begin{equation}
    \begin{split}
        \bullet\quad \rm{AIC} &= \chi ^{2} +2K,\\
        \bullet\quad \rm{BIC} &= \chi ^{2} +K\log(N_i),
    \end{split}
\end{equation}
where $\chi^{2}$ is calculated using the model's Gaussian likelihood function $\mathcal{L}(\hat{\theta} |data)$ value and $K$ is the number of free parameters for the particular model. At the same time, $N_i$ is the number of data points for the $i^{th}$ dataset. Accordingly, by defining
the AIC Bayes factor,
$\Delta \rm{\rm{AIC}} \equiv \rm{\rm{AIC}}_{\it f(Q)} - \rm{\rm{AIC}}_{\Lambda\rm{CDM}}$, when $\rm \Delta AIC \leq 2$   means that for the fitted data the proposed theoretical model holds a \textrm{substantial observational support}, $ 4 \leq {\rm \Delta AIC} \leq 7$  means  \textrm{less observational support}, and finally $\Delta \rm{AIC} \geq 10$ means \textrm{no observational support} as presented in \citep{szydlowski2015aic}, {all of them with respect to the $\Lambda$CDM corresponding benchmark.} {On the other hand, the BIC Bayes factor is defined as follows: \(\Delta \text{BIC}  \equiv 2\ln B_{ij} = -(\text{BIC}_{i} - \text{BIC}_{j}).\) For our case, ($i$) refers to the \(\Lambda\)CDM model, while ($j$) denotes the \(f(Q)\) gravity model.}  Thus, based again on the categorization in \cite{szydlowski2015aic}, the ranking of evidence against \(\Lambda\)CDM, i.e., in favor of the \(f(Q)\) gravity model,  is as follows: Negligible if \(0 \leq \Delta \rm{BIC} \leq 2\), positive if \(2 \leq \Delta \rm{BIC} \leq 6\), strong if \(6 \leq \Delta \rm{BIC} \leq 10\), and extremely strong if \(\Delta \rm{BIC} > 10\).

Then, the total $\chi^2$ for OHD+SNIa datasets 
and other statistical quantities such as the likelihood function ${\mathcal{L}(\hat{\theta}|data)}$, the Chi-Square is $\chi ^{2}$,  the reduced Chi-Square,  $\chi^{2}$-red, \rm{AIC}, $\Delta \rm{AIC}$, \rm{BIC} and $\Delta \rm{BIC}$ are presented in Table \ref{stastical1}  for OHD, SNIa and SNIa+OHD. 
\\
\\
{From our statistical results, the values of $\Delta \rm{AIC}$ for the $f(Q)$ gravity model read {$1.46$, $1.74$, and $1.42$} for the OHD, SNIa, and OHD + SNIa datasets, respectively, and the results show that the model is substantially observational supported according to this analysis. However, the corresponding results of $\Delta \rm{BIC}$ for this model yield {$3.21$, $7.20$, and $5.88$}, for OHD, SNIa, and OHD+SNIa datasets. The latter analysis implies that the power-law $f(Q)$ gravity model has positive support against the $\Lambda$CDM for the case of OHD and OHD+SNIa datasets. In contrast, when we use SNIa data alone, the model is strongly supported against $\Lambda$CDM\footnote{According to the work in \citet{parkinson2005testing,biesiada2007information}  due to the large number of data $N_i$ such as SNIa, AIC tends to favor models with more parameters. In contrast, BIC tends to penalize them.}. It is precisely due to this discrepancy in the competitiveness of this class of $f(Q)$ models as per the AIC and BIC criteria which impels us to consider the first-order perturbed equations of motion and to perform further investigations to compare the predictions of $f(Q)$ gravity power-law models with pertinent large-scale structure observational data.}
\section{Cosmological perturbations in $f(Q)$ gravity
}
\label{gravity1}
In this section, our main objective is to investigate the cosmological perturbations by applying the  {expressions obtained through the} 1+3 covariant formalism for $f(Q)$-gravity model. Thus, herein the growth structure $f(Q)$ gravity power-law models, the $\sigma_8+f$, $f\sigma_8$, $\sigma_8+f+f\sigma_8$ datasets. We also consider the joint analysis of growth structure data with cosmic measurements, namely: OHD+SNIa+$f+\sigma_8$, OHD+SNIa+$f\sigma_8$, OHD+SNIa+$f+\sigma_8+f\sigma_8$ datasets will be taken into account for further investigation. 
\\
\\
The general form of the Raychaudhuri equation \citep{castaneda2016some} can be written as 
\begin{equation}
	\dot{\theta}=-\frac{1}{3}\theta^2-\frac{1}{2}(\rho_{eff}+3p_{eff})+\nabla^a\dot{u}_a\;, \label{32}
\end{equation}
where $u_a$ is the four-vector velocity of the matter fluid. So, the Raychaudhuri equation for the case of the $f(Q)$ gravity model becomes the following.
\begin{eqnarray}
	&&\dot{\theta}=-\frac{1}{3}\theta^2-\frac{1}{2}(1+3w)\rho_m  +\rho_Q-\frac{\theta\dot{Q}}{2Q}\left(2QF''+F'\right)\nonumber\\&&  +\nabla^a\dot{u}_a \;,
\end{eqnarray}
where $w$ is the equation of state parameter for the matter fluid. The corresponding conservation equations for the effective fluid are given by
\begin{eqnarray}
	&&\dot{\rho}_{eff}+\theta(\rho_{eff}+p_{eff}) = 0\;,\\&&
	(\rho_{eff}+p_{eff})\dot{u}_a+\tilde{\nabla}_ap_{eff} = 0\;,	
\end{eqnarray} 
where 
$\tilde{\nabla}_a$ represents the covariant spatial gradient.  For the cosmological background level, we consider a homogeneous and isotropic expanding universe based on spatial gradients of gauge-invariant variables such as 
\begin{eqnarray}
	&&D^m_a=\frac{a}{\rho_m}\tilde{\nabla}_a\rho_m\;,\label{35} \qquad 
	Z_a=a\tilde{\nabla}_a\theta\;,\label{36}
\end{eqnarray}
representing the energy density and volume expansion of the fluid, respectively \citep{dunsby1992cosmological, abebe2012covariant} and which are the basic tools to extract the evolution equations for matter density fluctuations. Based on the non-metricity fluid for arbitrary $f(Q)$ gravity, we introduce the new terms $\mathcal{W}_a$ and $\mathcal{L}_a$, 
\begin{eqnarray}
	&&\mathcal{W}_a=a\tilde{\nabla}_aQ,\label{ggg} \qquad \mathcal{L}_a=a\tilde{\nabla}_a\dot{Q}\;,\label{oooo}
\end{eqnarray}
for the spatial gradients of gauge-invariant quantities to characterize the fluctuations in the nonmetricity density and momentum respectively.  By applying techniques similar to \citep{sami2021covariant}, the first-order linear evolution equations are derived from Eqs.  \eqref{35} - \eqref{oooo} as follows: 
\begin{eqnarray}
	&& \dot{D}^m_a=-(1+w)Z_a+w\theta D^m_a \;,\label{12} \\
	&&\dot{Z}_a=-\left[ \frac{2\theta}{3}+\frac{\dot{Q}F'}{2Q}-\dot{Q}F''\right]Z_a+\left[ \frac{w\theta^2}{3(1+w)}\nonumber\right.\\&&\left. -\frac{(1+3w)}{2(1+w)}\rho_m -\frac{wF}{2(1+w)}-\frac{QF'w}{1+w}+\frac{\theta\dot{Q}F''w}{1+w} \nonumber\right.\\&&\left. +\frac{\theta\dot{Q}F'w}{2Q(1+w)}- \frac{w}{1+w}\tilde{\nabla}^2\right] D^m_a+\left[ \frac{1}{2}F'   -QF'' \nonumber\right.\\&&\left. -F'+\theta\dot{Q}F'''-\frac{\theta\dot{Q}F''}{2Q}+\frac{\theta\dot{Q}F'}{2Q^2}\right]\mathcal{W}_a\nonumber\\&&  -\left[ \frac{\theta F'}{2Q}\mathcal -\theta F''\right] \mathcal{L}_a\;,\\&&
	\dot{\mathcal W}_a=\mathcal{L}-\frac{w \dot{Q}}{1+w}D^m_a\;,\\&&	
	\dot{\mathcal L}_a=\frac{\dddot{Q}}{\dot{Q}}\mathcal W-\frac{w \ddot{Q}}{1+w}D^m_a\;.\label{21}
\end{eqnarray}

 From Eqs. \eqref{12} - \eqref{21} we find the scalar perturbation equations, which are responsible for the formation of large-scale structures \citep{abebe2012covariant}.
To extract any scalar variable $Y$ from the first-order evolution equations, we perform the usual decomposition yielding
	\begin{equation}
		a\nabla_aY_b=Y_{ab}=\frac{1}{3}h_{ab}Y+\Sigma_{ab}^Y+Y_{[ab]}\;.
	\end{equation}
Here $Y=a\nabla_a Y^a$, whereas $\Sigma^Y_{ab}=Y_{(ab)}-\frac{1}{3}h_{ab}Y$ and $Y_{[ab]}$ 
represent the shear (distortion) and vorticity (rotation) of the density gradient field, respectively. Then, we define the following scalar quantities \citep{abebe2012covariant}
\begin{equation}
		\mathcal{D}_m=a\tilde{\nabla}^aD^m_a,\hspace{0.1cm}	Z=a\tilde{\nabla}^aZ_a,\hspace{0.1cm}\mathcal{W}=a\tilde{\nabla}^a\mathcal{W}_a,\hspace{0.1cm}\mathcal{L}=a\tilde{\nabla}^a\mathcal{L}_a.
	\end{equation}
The variables above evolve following first-order perturbation evolution equations as
\begin{eqnarray}
		&&\dot{\mathcal{D}}_m=-(1+w)Z+w\theta\mathcal{D}_m \label{5200} \;, \\
		&&\dot{Z}=-\left[ \frac{2\theta}{3}+\frac{\dot{Q}F'}{2Q}-\dot{Q}F''\right]Z+\left[ \frac{w\theta^2}{3(1+w)}+ \nonumber\right.\\&&\left. \frac{w(1+3w)}{2(1+w)}\rho_m -\frac{w F}{2(1+w)}-\frac{QF'w}{1+w}+\frac{\theta\dot{Q}F''w}{1+w} \nonumber\right.\\&&\left. +\frac{\theta\dot{Q}F'w}{2Q(1+w)}-\frac{1}{2}(1+3w)\rho_m   -\frac{w}{1+w} \tilde{\nabla}^2\right]\mathcal{D}_m+ \nonumber\\&& 
        \left[ \frac{1}{2}F'  -QF''-F'+\theta\dot{Q}F'''-\frac{\theta\dot{Q}F''}{2Q}   +\frac{\theta\dot{Q}F'}{2Q^2}\right]\mathcal{W}\nonumber\\&& -\left[ \frac{\theta F'}{2Q}\mathcal -\theta F''\right] \mathcal{L}\;,\label{577} \\&&
		\dot{\mathcal W}=\mathcal{L}-\frac{w \dot{Q}}{1+w}\mathcal{D}_m,\label{588}\\&&
		\dot{\mathcal L}=\frac{\dddot{Q}}{\dot{Q}}\mathcal W-\frac{w \ddot{Q}}{1+w}\mathcal{D}_m.\label{5100}
	\end{eqnarray}

After having found the system \eqref{5200} -\eqref{5100} of the scalar evolution equations, we apply the harmonic decomposition technique presented in \citep{abebe2012covariant, ntahompagaze2018study,sami2021covariant} to obtain the eigenfunctions with the corresponding wave number $\tilde{\nabla}^2  \equiv -{k^2}/{a^2}$ (where the wave number $k = \frac{2\pi a}{\lambda}$ \citep{dunsby1992cosmological} and $\lambda$ is the wavelength) for harmonic oscillator differential equations in $f(Q)$ gravity. To extract the eigenfunctions and wave numbers, the harmonic decomposition technique is applied to the first-order linear cosmological perturbation equations of scalar variables \citep{sahlu2020scalar} as those in \eqref{5200} -\eqref{5100}. For any second-order functions $X$ and $Y$ the harmonic oscillator equation is given as
	\begin{equation}
		\ddot{X}=A\dot{X}+BX-C(Y,\dot{Y} ),
	\end{equation}	
	where {the first, second and last terms on the right hand side of the equation represent the frictional, restoring, and source forces, respectively}, and the separation of variables takes the form
		$X=\sum_{k}X^k(t)Q^k(x), \hspace{0.1cm}{\rm and}  \hspace{0.1cm}	Y=\sum_{k}Y^k(t)Q^k(x)\;,$
	where $k$ is the wave number and $Q^k(x)$ is the eigenfunction of the covariantly defined Laplace-Beltrami operator in (almost) FLRW space-times, $
		\nabla^2Q^k(x)=-\frac{k^2}{a^2}Q^k(x).$
	 Then, the first-order evolution equations \eqref{5200} - \eqref{5100}  become:
\begin{eqnarray}
	&& \dot{\mathcal{D}}^k_m=-(1+w)Z^k+w\theta \mathcal{D}^k_m \;,\label{dog1123} \\
	&&\dot{Z}^k=-\Big[ \frac{2\theta}{3}+\frac{\dot{Q}F'}{2Q}-\dot{Q}F''\Big]Z^k+\Bigg[ \frac{w\theta^2}{3(1+w)}\nonumber\\&& -\frac{(1+3w)}{2(1+w)}\rho_m -\frac{wF}{2(1+w)}-\frac{QF'w}{1+w}+\frac{\theta\dot{Q}F''w}{1+w} \nonumber\\&&+\frac{\theta\dot{Q}F'w}{2Q(1+w)}+\frac{k^2}{a^2}  \frac{w}{1+w}\Bigg] \mathcal{D}^k_m  +\Bigg[ \frac{1}{2}F'  -QF'' \nonumber\\&&-F'+\theta\dot{Q}F'''-\frac{\theta\dot{Q}F''}{2Q}+\frac{\theta\dot{Q}F'}{2Q^2}\Bigg]\mathcal{W}^k \nonumber\\&& 
 -\left[ \frac{\theta F'}{2Q}\mathcal -\theta F''\right] \mathcal{L}^k\;,
 %
 \\&&
	\dot{\mathcal W}^k=\mathcal{L}^k-\frac{w \dot{Q}}{1+w}\mathcal{D}^k_m\;, \label{dog321}\\&&	
	\dot{\mathcal L}^k=\frac{\dddot{Q}}{\dot{Q}}\mathcal W^k-\frac{w \ddot{Q}}{1+w}\mathcal{D}^k_m\;.\label{dog321}
\end{eqnarray}
Consequently, the second-order evolution equations \eqref{dog1123} - \eqref{dog321}  with respect to time derivative yield as
	\begin{eqnarray}
		&&\ddot{\mathcal{D}}_m^k= -\left[ \frac{2\theta}{3}+\frac{\dot{Q}F'}{2Q}+\dot{Q}F''-w\theta\right]\dot{\mathcal{D}}_m^k-\left[-w F \nonumber\right.\\&&\left. -\theta\dot{Q}F''w  +\frac{\theta\dot{Q}F'w}{2Q}-\frac{(1+3w)\rho_m}{2}(1-w) \nonumber\right.\\&&\left. +\frac{k^2}{a^2} w\right]\mathcal{D}_m^k-\left[ \frac{1}{2}F'   -QF''-F'+\theta\dot{Q}F'''\nonumber\right.\\&&\left.-\frac{\theta\dot{Q}F''}{2Q}+\frac{\theta\dot{Q}F'}{2Q^2}\right](1+w)\mathcal{W}^k \label{70011}\nonumber\\&& +\left[ \frac{\theta F'}{2Q} -\theta F''\right](1+w)\dot{\mathcal{W}}^k\;, \\&&
		\ddot{\mathcal W}^k=\frac{\dddot{Q}}{\dot{Q}}\mathcal W^k-\frac{2w \ddot{Q}}{1+w}\mathcal{D}_m-\frac{w \dot{Q}}{1+w}\dot{\mathcal{D}}_m^k\,.\label{700x}
	\end{eqnarray}

 \subsection{Density contrast evolution}
 We shall consider the matter-dominated universe $w = 0$, and use the redshift-space transformation technique so that any first-order and second-order time derivative functions $\dot{Y}$ become
\begin{eqnarray}\label{transformation}
    &&\dot{Y} = -(1+z)HY'\;,\\&&
    \ddot{Y} = (1+z)H^2Y' +(1+z)^2H^2Y''+  (1+z)^2H'H Y'\;,\nonumber
\end{eqnarray}
to present the growth density fluctuations through cosmological redshift. 
Thus, by resorting to expressions {Eqs. \eqref{transformation}} and assuming the power law model $F(Q) = \alpha Q^n$, the second-order evolution equations Eqs. \eqref{70011} - \eqref{700x} read as follows:
\begin{eqnarray}
		&&(1+z)^2{\mathcal{D}}''_m= (1+z)\bigg\{ 1- (1+z)\frac{H'}{H} + \nonumber\\&& \frac{H'}{H}n\left({1-\Omega_m}\right)E^{2n-2}\bigg\}{\mathcal{D}}'_m 
  +\frac{3\Omega_m}{2E^2}(1+z)^3\mathcal{D}_m \nonumber\\&& -\frac{1}{H^2}\Bigg\{ \frac{(1-\Omega_m)}{(1-2n)}E^{2n-2}\bigg[\frac{n^2}{2}-n   -(1+z)\frac{{H'}}{H}n\bigg(n^2 \label{second1} \\&& -4n+\frac{7}{2}\bigg)\bigg] \Bigg\}\mathcal{W} -(1+z)\frac{n}{2H^2}\left(1-\Omega_m\right)E^{2n-2}{\mathcal{W}}'\;,\label{7001}\nonumber\\&&
		\left(1+z \right)^{2} \mathcal{W}'' =
    \frac{40n(\Omega_m+3-\Omega_m-9)}{3(2n-1)(n^2 \Omega_m-n^2+2n-1)}\mathcal{W}\;.\label{second2}~~~~~~~
	\end{eqnarray}
As we can see, Eqs. \eqref{70011} - \eqref{700x} are nonlinear evolution equations in which no approximation has been made. To convert such a set of equations into a closed second-order equation, many works \citep{sahlu2020scalar,sami2021covariant,Ntahompagaze,abebe2013large} have considered the quasi-static approximation. For simplicity, we applied the quasistatic approximation, where the first and second order of the time derivative of nonmetric density fluctuations is assumed to be approximately zero ($\dot{\mathcal{W}} = \ddot{\mathcal{W}} \approx 0$). Once the quasi-static approximation is made, Eqs. \eqref{70011} - \eqref{700x}  reduced to the closed system equation and it becomes  
\begin{eqnarray}
		&&\ddot{\mathcal{D}}_m^k= -\Bigg\{ \frac{2\theta}{3}+\frac{\dot{Q}F'}{2Q}+\dot{Q}F''-w\theta+\Bigg(\frac{1}{2}F'   -QF''\nonumber\\&&-F'+\theta\dot{Q}F''' -\frac{\theta\dot{Q}F''}{2Q}+\frac{\theta\dot{Q}F'}{2Q^2}\Bigg) \left(\frac{w \dot{Q}^2}{\dddot{Q}}\right) \Bigg\}\dot{\mathcal{D}}_m^k \nonumber\\&&
  -\Bigg\{\frac{\theta\dot{Q}F'w}{2Q}   -w F  -\theta\dot{Q}F''w  -\frac{(1+3w)\rho_m}{2}(1-w) \nonumber\\&&+\frac{k^2}{a^2} w  +\Bigg( \frac{1}{2}F'   -QF''  -F'+\theta\dot{Q}F'''-\frac{\theta\dot{Q}F''}{2Q} \nonumber\\&&  +\frac{\theta\dot{Q}F'}{2Q^2}\Bigg)\left(  \frac{2w \dot{Q}\ddot{Q}}{\dddot{Q}}\right) \Bigg\}\mathcal{D}^k_m \,. \label{70011x}
	\end{eqnarray}
For the case of \(n = 0\), Eqs.~\eqref{70011x} reduce to the usual \(\Lambda\)CDM equation for the density contrast as
\begin{eqnarray}
&&\ddot{\mathcal{D}}_m^k = -\left( \frac{2}{3} - w \right) \theta \dot{\mathcal{D}}_m^k 
+ \frac{1}{3}\theta^2 \Big[ 2w(1-\Omega_m) \nonumber\\&&
+ \frac{\Omega_m(1+3w)}{2}(1-w) 
- \frac{k^2}{3H^2a^2}w \Big] \mathcal{D}_m^k.
\label{125}
\end{eqnarray}
Once  redshift-space transformation is implemented, we find that 
quasi-static Eq. \eqref{70011x}, for dust ($\omega=0$) and $f(Q)$ power-law models \eqref{model},
yields 
\begin{eqnarray}
    && (1+z)^2{\mathcal{D}}''_m= (1+z)\Bigg[ 1- (1+z)\frac{H'}{H} + \\&& \,\frac{H'}{H}\left(1-\Omega_m\right)n E^{2n-2}\Bigg]{\mathcal{D}}'_m 
  +\,\frac{3\Omega_m}{2E^2}(1+z)^3\,\mathcal{D}_m\label{rr}\;.\nonumber
\end{eqnarray} 
%
To solve the density contrast numerically resorting the evolution equations \eqref{second1}-\eqref{second2}, we use the initial values of the system.  Due to gravitational instability, the variation in the temperature of the CMB is of the order $10^{-5}$ \citep{smoot1992structure} at $z_{in}\approx 1089$, and we consider it an initial condition. This small gravitational instability is the seed for the formation of large-scale structures.
Hence, we define the normalized energy density contrast for the matter fluid as
\begin{eqnarray}
	\delta(z)=\frac{\mathcal{D}_m(z)}{\mathcal{D}_m{(z_{in})}}\;,
\end{eqnarray}
where the subscript $in$ refers to the initial value of $\mathcal{D}_m (z)$ at the given initial redshift $z_{in} $ \citep{smoot1992structure}.  As stated previously, we consider two different situations in this section: the first illustration does not assume any (quasi-static) approximation (dubbed the full system in the following), whereas the second situation relies on the quasi-static approximation
to determine the evolution of the density contrast $\delta_m(z)$ in the matter-dominated universe. 
 \begin{figure*}
      \includegraphics[width=7.8cm, height=6cm]{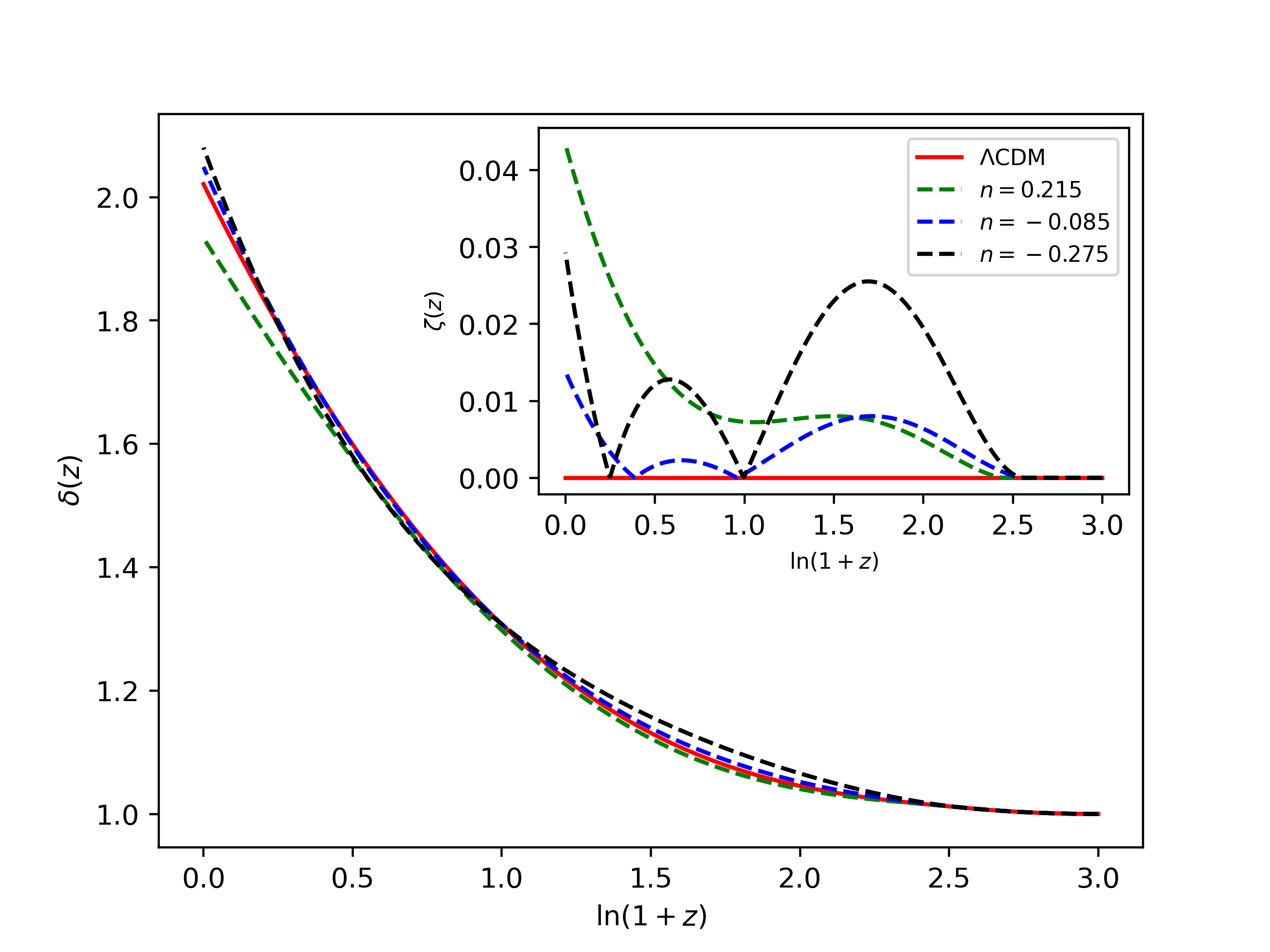 }
		\includegraphics[width= 7.8cm,height= 6cm ]{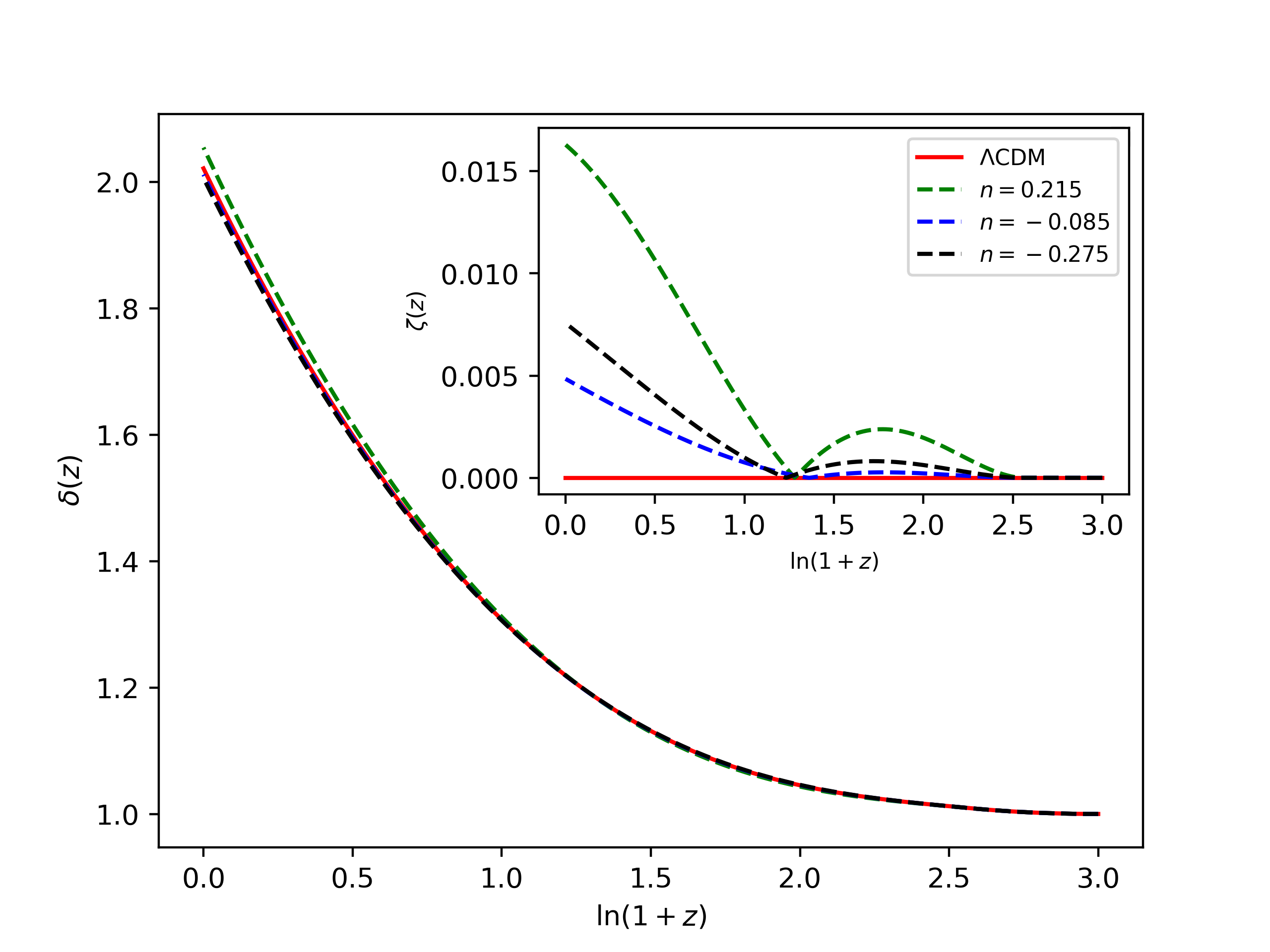}
		\caption{{{
  The outer left panel displays growth density fluctuations, $\delta(z)$ when resolving the whole system \eqref{second1}-\eqref{second2}, while the inner left panel illustrates the relative difference, $\zeta(z)$, between the $\Lambda$CDM and $f(Q)$-power-law models across the cosmological redshift. The outer right panel depicts growth density fluctuations $\delta(z)$ under the quasi-static approximation given by \eqref{rr}, and the inner right panel showcases the relative difference $\zeta(z)$, between the $\Lambda$CDM and $f(Q)$ power-law models. These plots correspond to values of {$n  = -0.085^{+0.300}_{-0.190} = \{0.215,-0.085,-0.275\}$ and $\Omega_m = 0.304$}, which are obtained from the best-fit parameters derived using OHD + SNIa data (see Table \ref{Tableone}).}}}\label{fig:q1}
\end{figure*}
\subsection{Numerical results of $\delta(z)$ } 
We first consider the evolution equation without any approximation as presented in Eqs. \eqref{second1} - \eqref{second2}, and show the results for the density contrast in the left outer panel of Fig. \ref{fig:q1}.
using the best-fit values of $n$.  Second, we have studied the growth of density fluctuations resorting to the quasi-static approximation as per Eq. \eqref{rr} whose results are presented in the right outer panel of Fig. \ref{fig:q1} for the matter-dominated universe. 
Meanwhile, in order to represent the density contrast $\delta(z)$ for the model $f(Q)$, for illustrative purposes, we have chosen the best-fit values of {$\Omega_m =0.304$ and  $n  = -0.085^{+0.300}_{-0.190} = \{0.215,-0.085,-0.275\}$}  as obtained from the analysis of OHD + SNIa presented in Table \ref{Tableone}. For $\Lambda$CDM we have also chosen $\Omega_m = 0.323$ so a comparison becomes possible. 

The numerical values of the fluctuation of matter density $\delta(z)$ turned out to be very sensitive to the constrained values of $\Omega_m$ and the exponent $n$ in both scenarios (full and quasi-static).  We have also presented the relative difference of $\delta(z)$ between $\Lambda$CDM and $f(Q)$ model by defining the dimensionless parameter $\zeta(z)$ as
\begin{eqnarray} \label{varaince}
    \zeta (z) = \bigg| \frac{\delta^{\rm \Lambda CDM}(z)-\delta^{f(Q)}_{\rm {full,QSA}}(z)}{\delta^{\rm \Lambda CDM}(z)}\bigg|\;\;,
\end{eqnarray}
where $\delta^{f(Q)}_{\rm {full, QSA}}$ refer to the $f(Q)$-gravity model density contrast as obtained for either the full system or quasi-static (QSA) approximation, respectively. The numerical results of $\zeta(z)$ are presented in the inner left and right panels of Fig. \ref{fig:q1}.
\begin{figure}
\centering
 \includegraphics[width= 8cm,height= 5cm ]{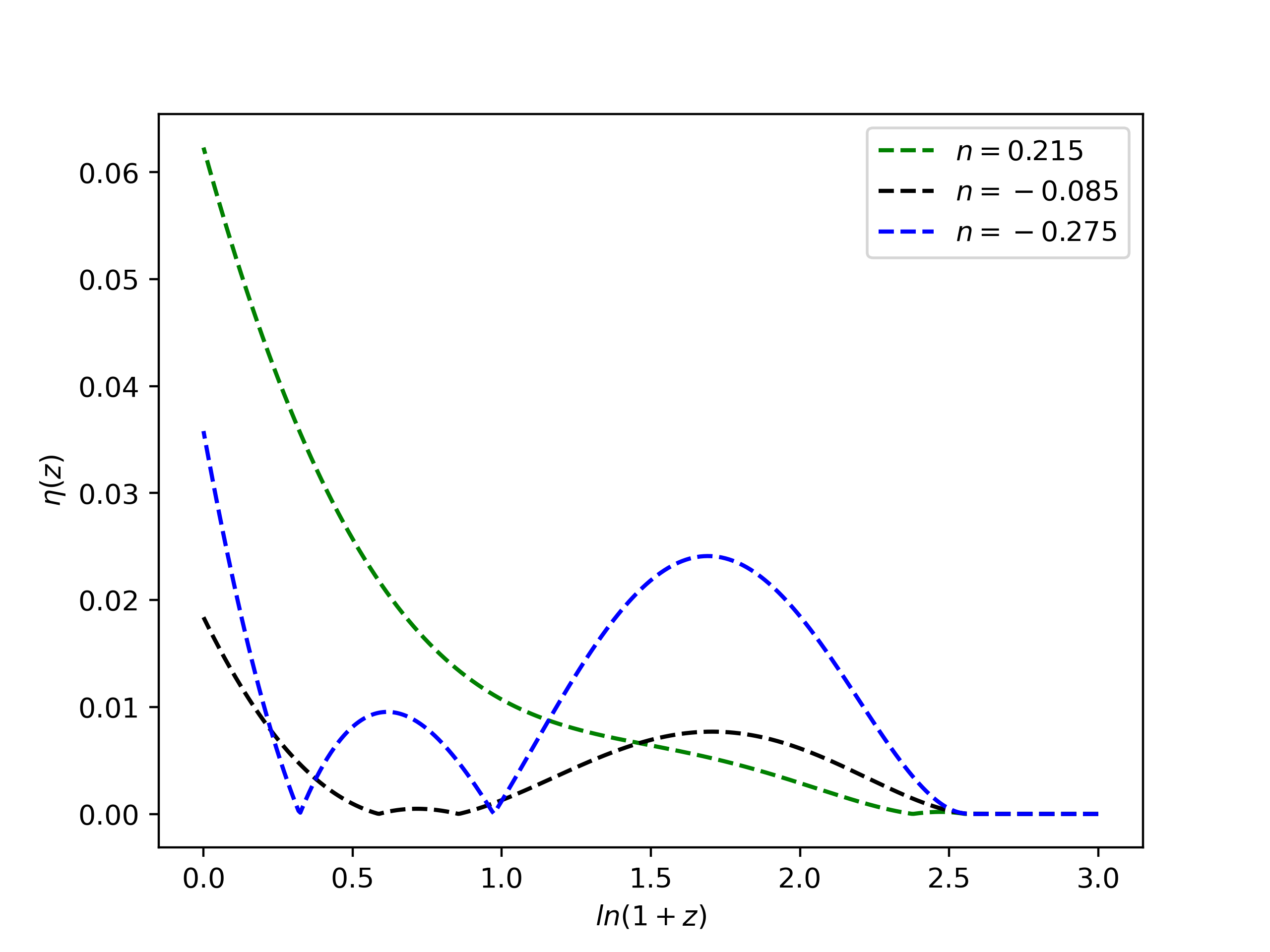}
		\caption{The relative difference between the full system and the quasi-static resolutions for $f(Q)$-gravity model for the values of  {$n  = -0.085^{+0.300}_{-0.190} = \{0.215,-0.085,-0.275\}$and $\Omega_m = 0.304$} which have been taken from the best-fit values obtained using OHD+SNIa data, see Table \ref{Tableone}.}
		\label{fig:qqxx}
\end{figure}
Complementary, to compare the quasi-static findings with the full system results, we introduce the dimensionless parameter   
\begin{eqnarray}\label{relative}
    \eta(z) = \bigg|\frac{\delta
    ^{f(Q)}_{\rm {QSA}}(z)-\delta^{f(Q)}_{\rm {full}}(z)}{\delta^{f(Q)}_{\rm {full}}(z)}\bigg|\;,
\end{eqnarray}
and presented its numerical finding in Fig. \ref{fig:qqxx}. This definition serves as an indicator of the competitiveness of quasi-static with respect to the full system resolution. From Fig. \ref{fig:qqxx}, we conclude that the quasi-static approximation better represents the evolution of the density contrast for values of $n$ closer to zero ($n=0$ corresponding to $\Lambda$CDM). Based on this result, we highlight that the quasi-static approximation, which has been widely taken into consideration in other modified theories of gravity \citep{ abebe2012covariant,sami2021covariant, abebe2015breaking, Bos, Sawicki}  may be used to represent with a competitive {94.03\% at $n = 0.215$, 98.7\% at $n = -0.085$, and 96.61\% at $n = -0.275$}, at redshift $z=0$ where the solutions differ the most, degree of accuracy, growth of structures in the context of $f(Q)$ power-law models. Consequently, this allowed us to rely on this approximation to analyse structure growth as presented in Sec. \ref{sigma}. 

\section{Structure growth in $f(Q)$ cosmology}
\label{sigma}
In this section, we study the ability of the $f(Q)$-gravity power-law models to fit large-scale structure data statistically. In order to do so, the linear growth rate $f$, as obtained from the density contrast $\mathcal{D}_m$, yields 
\begin{equation}
    f 
    \equiv       
    \frac{{\rm d}\ln{\mathcal{{D}}_m}}{{\rm d}\ln{a}} = -(1+z)\frac{1}{\mathcal{D}_m} \frac{{\rm d}\mathcal{D}_m(z)}{{\rm d}z}    
    \;.\label{growth1}
\end{equation}
Thus, by substituting the definition of \eqref{growth1} into the quasi-static second-order evolution equation \eqref{rr}, for the matter-dominated universe, the evolution of the growth rate is governed by the following expression\footnote{For the case of $n = 0$, Eq. \eqref{growthrate} exactly reduced to $\Lambda$CDM. Indeed, by resorting to Eqs. \eqref{333} and \eqref{transformation} it is straightforward to obtain
\begin{eqnarray}\label{growthratelcdm}
    && (1+z)f' = f^2 -\left[(1+z)\frac{H'}{H}-2\right] f -\frac{3 \Omega_m}{2E^2}(1+z)^3 \label{rr1}\;. \nonumber\\&&
\end{eqnarray}}
\begin{eqnarray}\label{growthrate}
     &&(1+z)f' = f^2 -\left[ (1+z)\frac{H'}{H}  +\left(1-\Omega_m\right)n E^{2n-2}-2\right]f \nonumber\\&&
  -\,\frac{3\,\Omega_m(1+z)^3}{2\,\left[\Omega_m(1+z)^3+(1-\Omega_m)E^{2n}\right]}\label{rr1}\;.
\end{eqnarray}  
A combination of the linear growth rate $f(z)$ with the root mean square normalization of the matter power spectrum $\sigma_8$ within the radius sphere $8h^{-1}$Mpc, yields the redshift-space distortion $f\sigma_8$ as
\begin{eqnarray}\label{growth11}
  f\sigma_8(z)  = -(1+z)\sigma_8\frac{\mathcal{D}'_m(z)}{\mathcal{D}_m(z)}\;.
 \end{eqnarray}
%
In the following two consecutive subsections, we characterize and constrain the effect of our
$f(Q)$-gravity model in the growth of structures, by admitting that the sets of observables $\sigma_8+f$, $f\sigma_8$ and $\sigma_8+f+f\sigma_8$ - as explained in the Introduction - are all that we have available to compare with theoretical predictions issued from equations \eqref{growthrate} and \eqref{growth11}.
 \begin{figure*}
    \includegraphics[width=.43\textwidth]{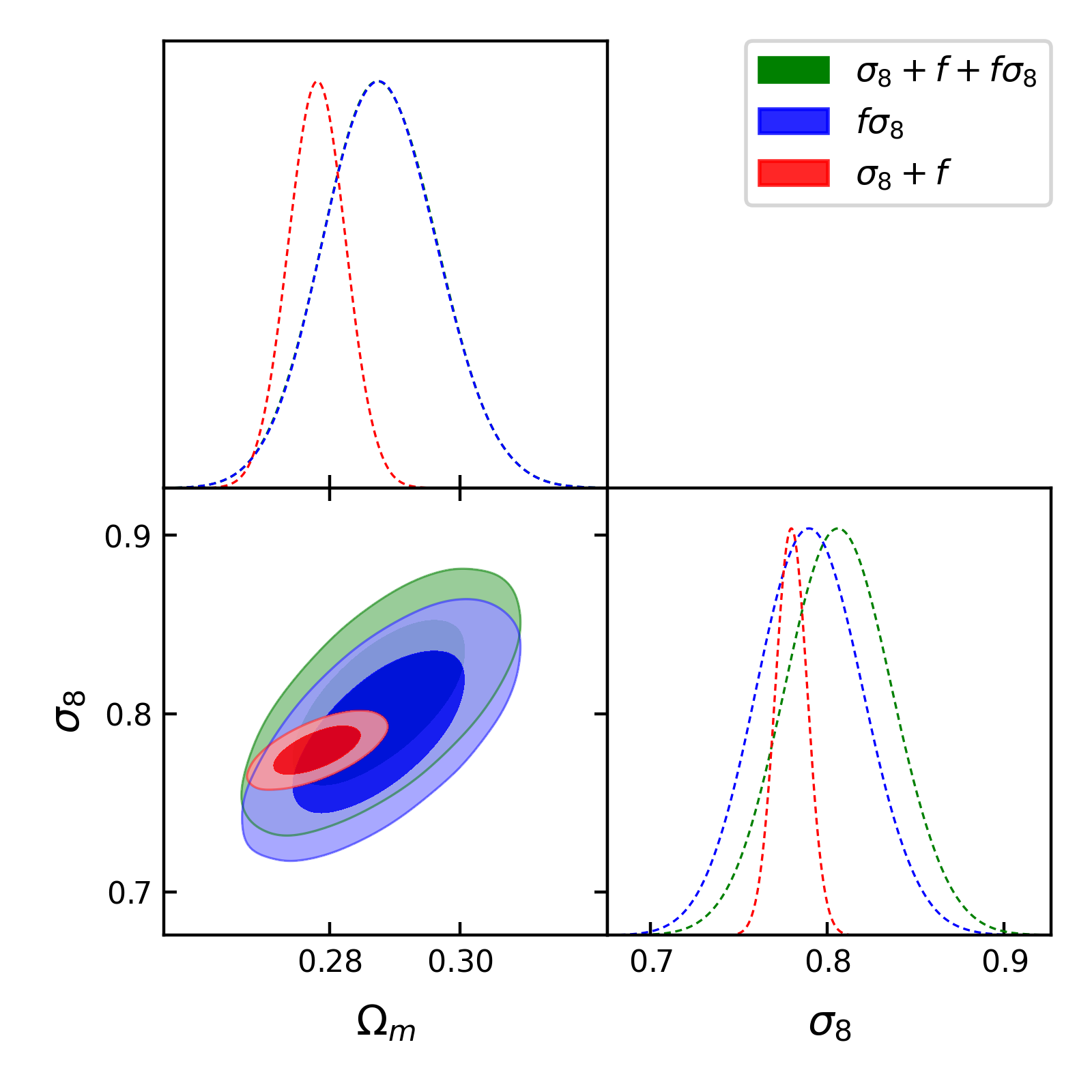}
    \hfill
    \includegraphics[width=.43\textwidth]{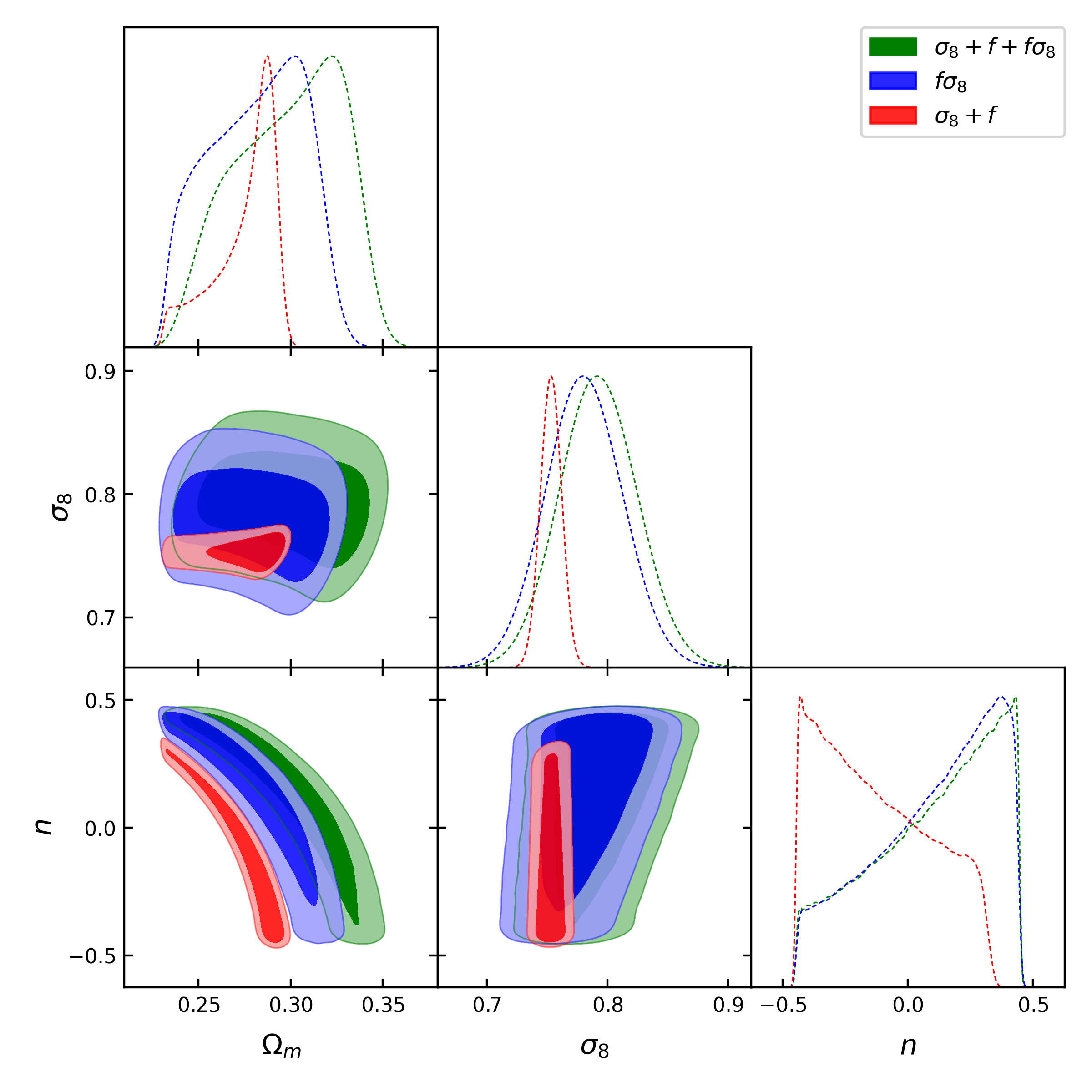}
    \caption{Joint MCMC contour plots for the best-fit values of the cosmological parameters $\Omega_m$, $\sigma_8$, and the exponent $n$ for both the $\Lambda$CDM model (left panel) and the $f(Q)$ gravity power-law model (right panel), using three datasets: $\sigma_8+f$, $f\sigma_8$, and $\sigma_8+f+f\sigma_8$. The use of $\sigma_8+f$ data seems to reduce the contour plots area notably for both classes of theories.}
    \label{fig:enter-label}
\end{figure*}
\begin{figure*}
 \includegraphics[width=.45\textwidth]{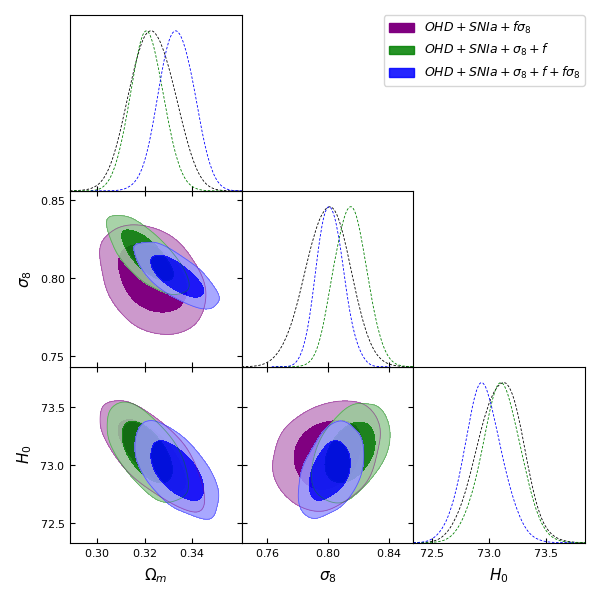}
 \includegraphics[width=.45\textwidth]{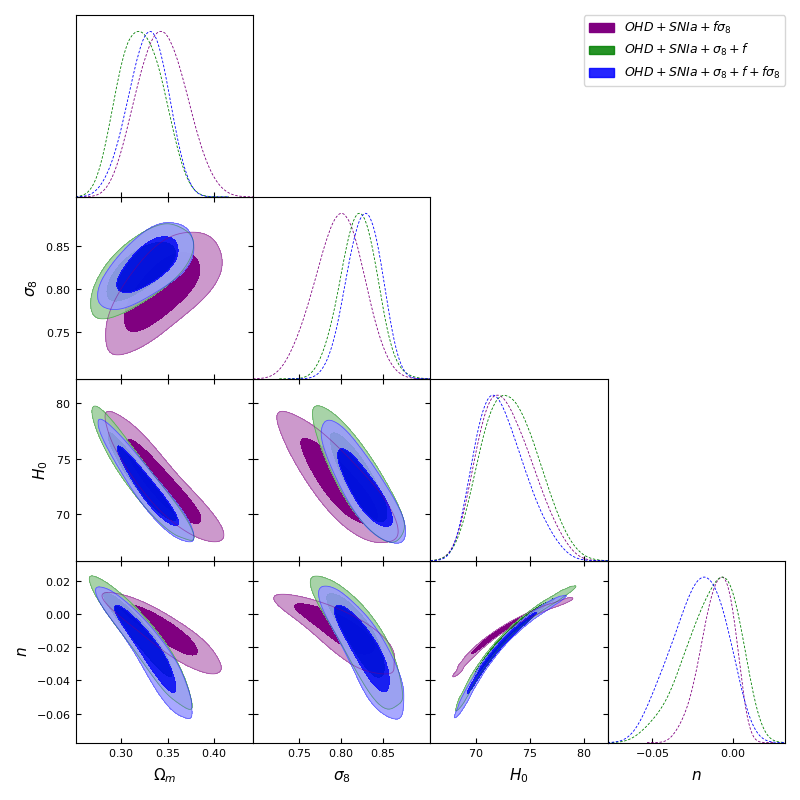}
	  \caption{{The constrained parameters of $\Omega_m$, $\sigma_8$, and $H_0$ for the $\Lambda$CDM model (left panel) and $\Omega_m$, $\sigma_8$, $H_0$, and $n$ for the $f(Q)$ gravity model (right panel). We used large-scale structure data - $f\sigma_8$, $\sigma_8+f$ and $\sigma_8+f+f\sigma_8$ - together with OHD + SNIa datasets to refine the best-fit values of these parameters. The use of $\sigma_8+f$ data seems to reduce the contour plots area notably for both classes of theories.} }
   
	 \label{ohdsnfsigma81}
	 \end{figure*}
     \begin{figure}
 \includegraphics[width= 8cm,height= 6cm]{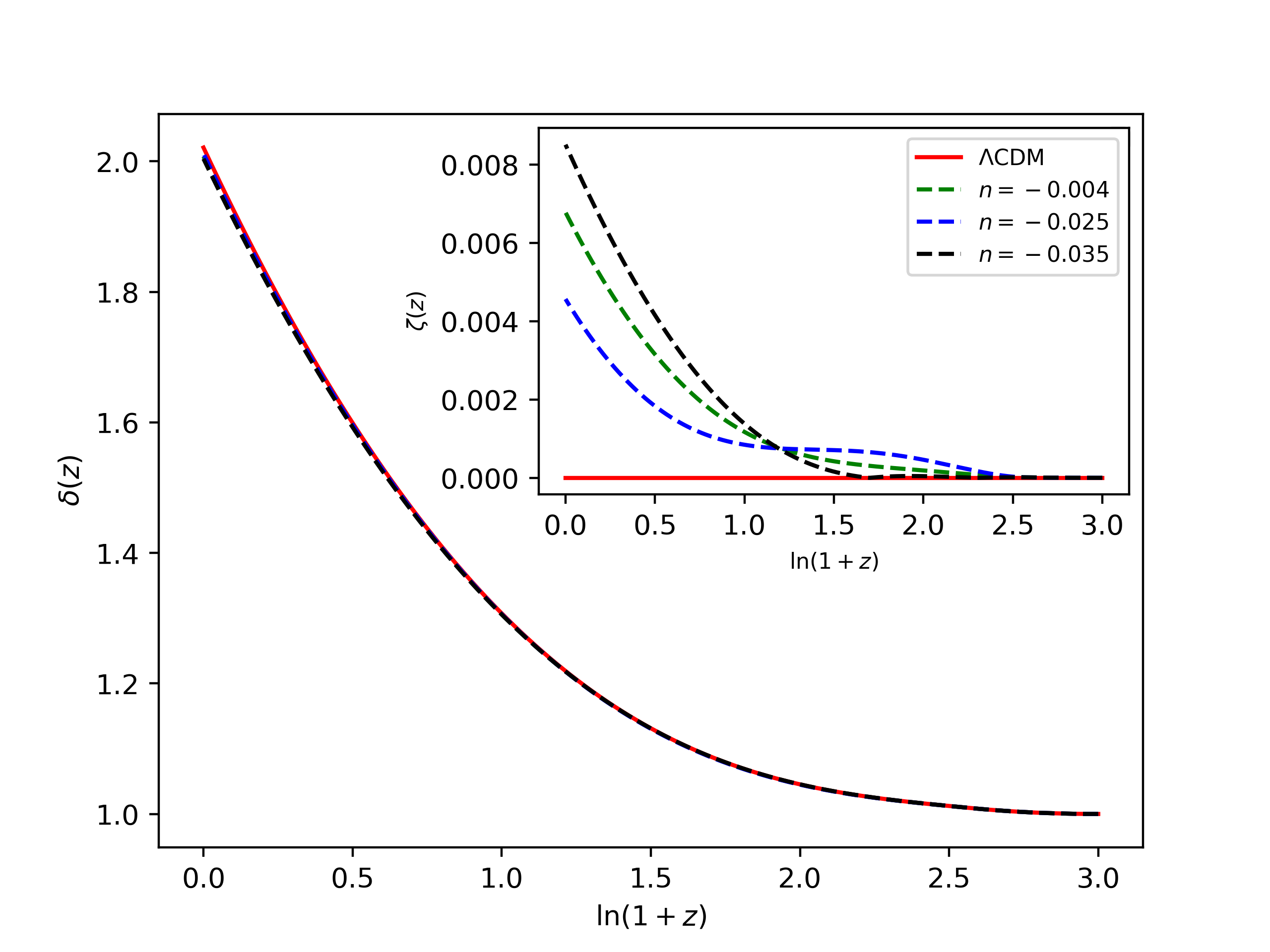}
	  \caption{Outer-Panel: Growth of density fluctuations for the matter-dominated universe as per Eq. \eqref{rr}, i.e., within the quasi-static approximation. Inner panel: redshift evolution for the deviation $\zeta(z)$ between $\Lambda$CDM and $f(Q)$-gravity model as per Eq. \eqref{varaince}.  For illustrative purposes, we use the best-fit values of $\Omega_m = 0.327$, and $n =-0.025^{+0.021}_{-0.010} = \{-0.004, -0.025, -0.035\}$ obtained from OHD + SNIa + $\sigma_8+f\,+\,f\sigma_8$ datasets, see Table \ref{bestfitparameters}.} 
	 \label{fig:matter-universe}
     \end{figure}
     \begin{figure}
   \includegraphics[width= 8cm,height= 6cm]{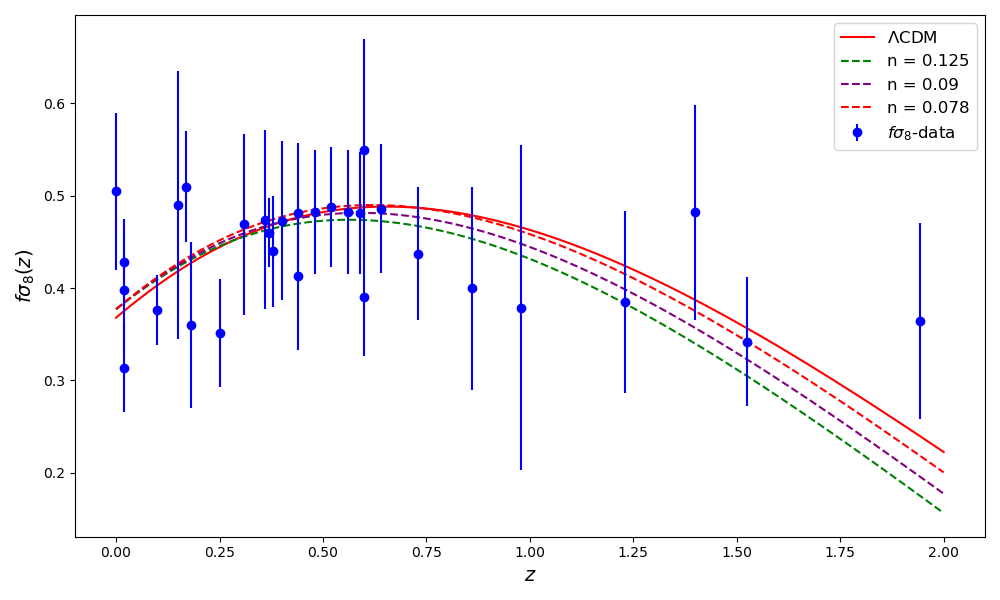}
	  \caption{{Linear growth-structure $f\sigma_8$ diagrams for $\Lambda$CDM and  $f(Q)$-gravity power-law models  - for different values of $n$ - with $f\sigma_8$ data. For illustrative purposes, we use $n = 0.090^{+0.035}_{-0.012} = \{0.125, 0.09, 0.078 \}$ with the corresponding values of $\sigma_{8} = 0.792^{+0.012}_{-0.012}$ and $\Omega_{m} = 0.300^{+0.037}_{-0.022}$ as shown in Table \ref{bestfitparameters}. For the latter two parameters, $\sigma_8$ and $\Omega_m$- we have taken the central values.
   }}
	 \label{fig:growth}
  \end{figure}  
\begin{figure*}
    \includegraphics[width=1.\linewidth]{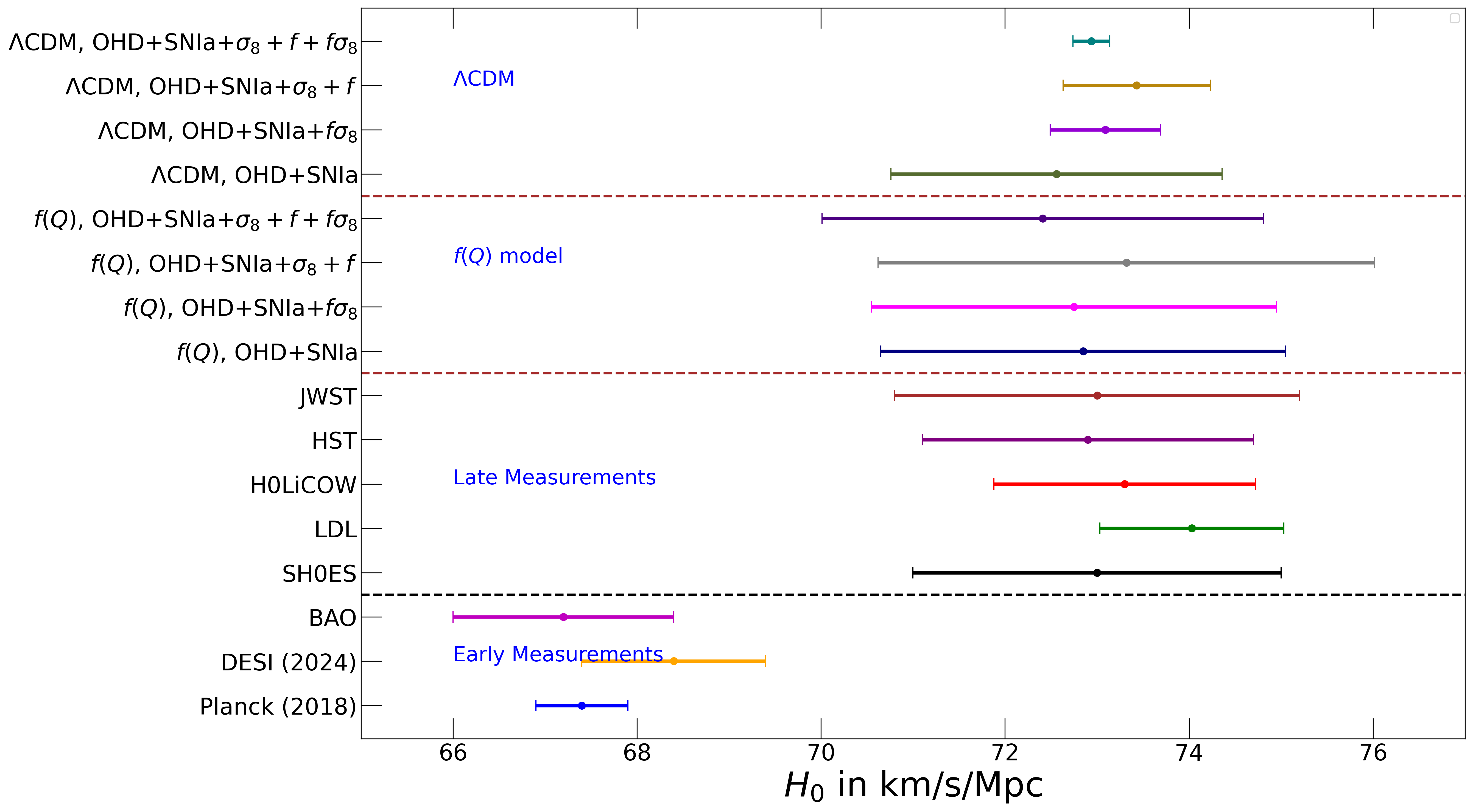}
    \caption{{Comparison of $\Lambda$CDM and $f(Q)$-gravity model $H_0$ values (in km/s/Mpc) with various measurements.}}
    \label{fig:enter-labelH0}
    \includegraphics[width = 1.\linewidth]{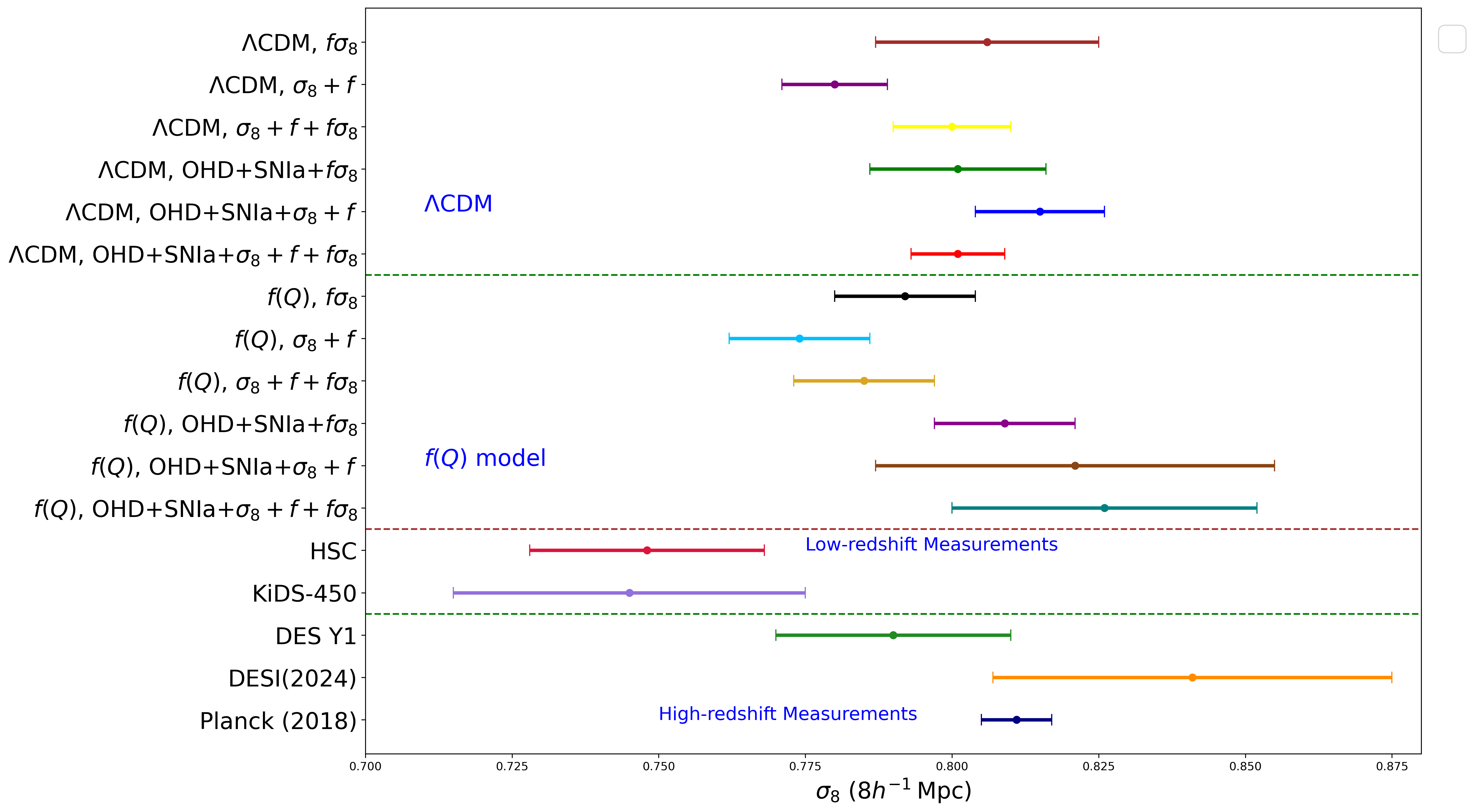}
    \caption{{Comparison of $\Lambda$CDM and $f(Q)$-gravity model $\sigma_8$ values (in 8$h^{-1}$\,Mpc) with various measurements.}}
    \label{fig:enter-labelS8}
\end{figure*}

\begin{table*}
		\begin{tabular}{|l|l|l|l|l|l}
			\hline
			Parameters & $\Omega_m$ & $\sigma_{8}$ &  $n$ & $H_0$  \\
            &&&&$(\rm km\,s^{-1}Mpc^{-1})$\\
            \hline
            $\Lambda$CDM&&&&\\
            $f\sigma_8$&  $0.288^{+0.023}_{-0.022}$ & $0.806^{+0.019}_{-0.018}$ &- & -\\
          $\sigma_8+f$	&   $0.278^{+0.004}_{0.004}$ & $0.780^{+0.009}_{0.009}$ &-&-\\
           $\sigma_8 + f + f\sigma_8$& $0.290^{+0.009}_{0.009}$& $0.790^{+0.013}_{-0.013}$&-&-\\
   OHD + SNIa + $f\sigma_8$& {$0.323^{+0.009}_{-0.009}$}& {$0.800^{+0.015}_{-0.015}$}&- &{$73.09^{+0.60}_{-0.60}$}\\
   OHD + SNIa + $\sigma_8+f$& {$0.321^{+0.007}_{-0.007}$}& {$0.815^{+0.011}_{-0.011}$}&- &{$73.43^{+0.82}_{-0.75}$} \\
   OHD + SNIa + $\sigma_8+f+f\sigma_8$& {$0.336^{+0.007}_{-0.007}$}& {$0.801^{+0.008}_{-0.008}$}&- &{$72.94^{+0.17}_{-0.19}$}\\
			\hline
   $f(Q)$ model&&&&\\
     $f\sigma_8$  & $0.300^{+0.037}_{-0.022}$& $0.792^{+0.012}_{-0.012}$& $0.090^{+0.035}_{-0.012}$ &- \\        
          $\sigma_8+f$  &$0.275^{+0.020}_{-0.065}$&$0.774^{+0.009}_{0.009}$&$-0.120^{+0.013}_{-0.011}$ &-\\
          $\sigma_8+f+f\sigma_8$  &$0.282^{+0.033}_{-0.021}$&$0.785^{+0.012}_{-0.012}$&$0.100^{+0.034}_{-0.011}$&- \\
OHD + SNIa + $f\sigma_8$&{$0.335^{+0.058}_{-0.051}$}&{$0.809^{+0.012}_{-0.019}$}&{$-0.01^{+0.019}_{-0.018}$}&{$72.75^{+2.20}_{-2.50}$}\\
OHD + SNIa + $\sigma_8+f$&{$0.320^{+0.047}_{-0.039}$}&{$0.821^{+0.034}_{-0.034}$}&{$-0.013^{+0.010}_{-0.009}$}&{$73.32^{+2.70}_{-3.90}$}\\
          OHD + SNIa + $\sigma_8+f+f\sigma_8$&{$0.327^{+0.056}_{-0.045}$}&{$0.826^{+0.026}_{-0.037}$}&{$-0.025^{+0.021}_{-0.010}$}&{$72.41^{+2.40}_{-3.10}$}\\
            \hline
	\end{tabular}
	\caption{{{Best-fit values for cosmological parameters $\sigma_{8}$, $\Omega_{m}$, $H_0$ and $n$ for both $\Lambda$CDM and $f(Q)$-gravity power-law models using $f\sigma{_8}$, $\sigma_8+f$, $\sigma_8+f\,+\,f\sigma_8$, OHD + SNIa + $f\sigma_8$, OHD + SNIa + $\sigma_8+f$ and OHD + SNIa + $\sigma_8+f\,+\,f\sigma_8$. The CMB Planck-2018 value for the matter fluctuation amplitude ($\sigma_8)$ is $\sigma_8 = 0.811\pm 0.006$ \citep{aghanim2020planck}.}}
 }
 \label{bestfitparameters}
\end{table*}

\subsection{Constraining parameters}
In this section, we take into account the growth structure data from Table \ref{datasetsofgrowth1} and \ref{datasetsofgrowth} (see Appendix \ref{Appendix}), $\sigma_8+f$, $f\sigma_8$ and $\sigma_8+f+f\sigma_8$ to obtain the best-fit values of $\Omega_m$, $\sigma_8$, and $n$ through the MCMC simulations. These best-fit values of these parameters are also provided using the combinations of cosmic expansion measurements with the growth structure data, meaning that the OHD + SNIa +$f\sigma_8$, OHD + SNIa +$\sigma_8+f$, and OHD + SNIa +$\sigma_8+f+f\sigma_8$ datasets for both the $\Lambda$CDM model and the $f(Q)$ gravity power-law models.  All constrained parameters are presented in Table \ref{bestfitparameters}. 
%
Furthermore, in Fig. \ref{fig:enter-label} we provide the combined analysis of the constrained parameters using these three datasets $\sigma_8+f$, $f\sigma_8$, and $f\sigma_8+f+\sigma_8$, whereas in  Fig. \ref{ohdsnfsigma81}
we present the contour plots obtained when those growth data are used together with cosmic expansion measurements (i.e., OHD + SNIa + $f\sigma_8$, OHD + SNIa + $\sigma_8+f$ and OHD + SNIa + $\sigma_8+f\,+\,f\sigma_8$). In both figures \ref{fig:enter-label} and \ref{ohdsnfsigma81},  $1\sigma$ and $2\sigma$ levels are depicted, and the corresponding $\Lambda$CDM contour plots appear on the left panels for comparison. A  simple inspection of those two figures leads us to conclude that the consideration of the growth rate $f$ and the amplitude matter fluctuations $\sigma_8$ measurements from VIPERS and SDSS only (Fig. \ref{fig:enter-label}) or combined with OHD + SNIa (Fig. \ref{ohdsnfsigma81}), significantly improves the constraints on the parameters of the model. Nonetheless, when standard redshift-space distortion data are considered too, such an improvement is reduced, although still present, particularly when  OHD + SNIa are considered (Fig. \ref{ohdsnfsigma81}, {green and blue} contours remain of similar shape and area). 
Finally, by considering the best-fit parameters of {$\Omega_m = 0.327$, and $n =-0.025^{+0.021}_{-0.010} = \{-0.004, -0.025, -0.035\}$} as obtained from the fitting against OHD + SNIa + $\sigma_8+f\,+\,f\sigma_8$ data (see Table \ref{bestfitparameters}), we have plotted in Fig. \ref{fig:matter-universe} the evolution of density contrast $\delta(z)$ as governed by Eq. \eqref{rr}.

In addition to constraining the best-fit values of the parameters, in Fig. \ref{fig:growth} we have also presented the redshift-space distortion diagram $f\sigma_8$  using the evolution - quasi-static- equation Eq. \eqref{rr} for the matter-dominated universe.  Therein, the redshift-space distortion best fittings are shown for both $\Lambda$CDM and $f(Q)$-gravity power-law models. To present the $f\sigma_8$ diagram, we consider the best-fit values $\sigma_{8} = 0.806$, $\Omega_{m} = 0.288$ for $\Lambda$CDM as given in Table \ref{bestfitparameters}. For the case of the $f(Q)$ gravity model, we have used different values of  $n = 0.090^{+0.035}_{-0.012} = \{0.125, 0.090, 0.078 \}$ with the corresponding values of  $\sigma_{8} = 0.792^{+0.012}_{-0.012}$ and $\Omega_{m} = 0.300^{+0.037}_{-0.022}$. For the two latter parameters - $\sigma_8$ and $\Omega_m$ - we have taken the central values. All these values are taken from the results presented in Table \ref{bestfitparameters}. {
For both $\Lambda$CDM and $f(Q)$-gravity models,
in Fig. \ref{fig:enter-labelH0}, we compare the $H_0$ values as obtained from the joint analyses as shown in Table \ref{bestfitparameters}  with different current measurements. The comparisons are made resorting to $i)$ early measurements, including Planck 2018 \citep{aghanim2020planck}, Dark Energy Spectroscopy Instrument (DESI) \citep{adame2024desi}; and $ii)$ late-time measurements derived from direct observations of the local Universe, such as Supernovae and $H_0$ for the Equation of State (SH0ES) \citep{riess2019large}, H0LiCOW \citep{wong2020h0licow}, and Hubble Space Telescope (HST) \citep{riess20113}. Figure \ref{fig:enter-labelH0} shows the discrepancy of $H_0$ measurements from early-universe evidence, late-universe observations, and the theoretical models \(\Lambda\)CDM and \(f(Q) \) studied herein. The usual \(\Lambda\)CDM model corresponds well with the late measurements, which predict a higher \(H_0 \), while late-time measurements from sources like SH0ES and JWST consistently provide larger values. {The \(f(Q) \) model yields higher \(H_0 \) values, suggesting a better agreement with late observations, while it does not fully resolve the discrepancy of $H_0$ measurements.}}

Analogously, the comparison of  \(\sigma_8\) results with different experimental findings is showcased in Fig. \ref{fig:enter-labelS8} resorting to $i)$ low-redshift measurements ($z\leq1$) namely HSC \citep{hamana2020cosmological}, and KiDS-450 \citep{hildebrandt2017kids}; and $ii)$ high-redshift measurements ($z>1$) namely DESI \citep{adame2024desi}, and Planck 2018 \citep{aghanim2020planck}. {Such a figure shows that, with the exception of the $\sigma_8+f$ datasets, the best-fit values of $\sigma_8$ found with our analyses remain close to the high-redshift measurements for both $\Lambda$CDM and $f(Q)$ gravity models.}

\begin{table*}
		\begin{tabular}{|llllllll|}
			\hline
			{\bf Data} & \textbf{$\mathcal{L}(\hat{\theta}|{\rm data})$} & \textbf{$\chi ^{2}$} &  $\chi ^{2}$-red & \textbf{$\rm{AIC}$} &\textbf{$\Delta \rm{AIC}$} & \textbf{$\rm{BIC}$} & \textbf{$\Delta \rm{BIC}$}\\
   \hline
   $\Lambda$CDM&&&&&&&\\
   &&&&&&&\\
            $f\sigma_8$& -8.00& 16.00&0.57&20&--&24.80&--\\
          $\sigma_8+f$	  & -1.20 &  2.40  & 0.60&6.40 & --& 5.98 & --\\
             $\sigma_8+f+f\sigma_8$ &-9.10& 18.20&0.54&22.20&--&25.60&--\\
             OHD+SNIa+$f\sigma_8$  &-791.24& 1582.48&0.93&1588.48&--&1605.93&--\\
             {OHD+SNIa+$\sigma_8+f$}  &-783.83& 1567.66&0.87&1573.66&--&1590.04&--\\
             OHD+SNIa+$\sigma_8+f+f\sigma_8$  &-793.89& 1587.78&0.91&1593.78&--&1610.23&--\\
			\hline
		$f(Q)$ model&&&&&&&\\
     &&&&&&&\\
            $f\sigma_8$ &-8.90& 17.80&0.75&23.80&3.80&28.00&3.20\\
          $\sigma_8+f$	  & -1.82 & 3.64 & 0.91&9.64 & 3.24 & 7.00&1.21 \\
             $\sigma_8+f+f\sigma_8$ &-9.04& 18.08&0.60&24.08&1.88 &28.46&3.04 \\
             OHD+SNIa+$f\sigma_8$  &-790.15& 1581.04&0.92&1589.04&0.56&1610.96&5.03\\
             OHD+SNIa+$\sigma_8+f$  &-783.05& 1566.10&0.86&1574.1& 0.45&1595.96&5.92\\
             OHD+SNIa+$\sigma_8+f+f\sigma_8$  &-793.450& 1586.90&0.89&1594.90&1.12&1616.06&5.82\\
			\hline
	\end{tabular}
	\caption{
 %
{
Values of ${\mathcal{L}(\hat{\theta}|data)}$, $\chi^{2}$, $\chi ^{2}$-red,  $\rm{AIC}$, $\Delta \rm{AIC}$, $\rm{BIC}$ and $\Delta \rm{BIC}$ for $\Lambda$CDM and $f(Q)$-gravity models. The $\Delta \rm{AIC}$ values are lower than $3.80$ for all datasets, whereas the $\Delta \rm{BIC}$ values have been found to lie within the range of observational support, {except for the case of OHD+SNIa+$\sigma_8+f$, the latter indicating less observational support. Interestingly, the OHD+SNIa+$f\sigma_8$ analysis indicates strong support when resorting to both criteria. Then, the addition of just three $\sigma_8+f$  data to OHD+SNIa+$f\sigma_8$ reduces the support, as per $\Delta$BIC, significantly although according to $\Delta$AIC, the dataset OHD+SNIa+$\sigma_8+f + f\sigma_8$  still remains strongly supported.}}
 \label{Tab:values}}
\end{table*}

\subsection{Statistical analysis }
Another crucial point in this section is to present the statistical values of ${\mathcal{L}(\hat{\theta}|{\rm data})}$, reduced $\chi^{2}$, $\chi ^{2}$, $\rm{AIC}$, $\Delta \rm{AIC}$, $\rm{BIC}$, and $\Delta \rm{BIC}$ to establish the viability of the $f(Q)$ power-law model. As mentioned above, the $f(Q)$ gravity models will be tested against the $\Lambda$CDM model, which will be considered as the benchmark accepted model. In order to do so, the growth structure data from Tables \ref{datasetsofgrowth1} and \ref{datasetsofgrowth}, as well as the cosmic expansion measurements from Section \ref{gravity} have been considered for a comprehensive analysis. 
{
From the results presented in Table \ref{Tab:values} we conclude that the $f(Q)$ power-law model has observational support based on our model selection criteria. In particular, $\Delta$AIC values lie between 0.34 and 3.80 for  $f\sigma_8$ and $\sigma_8+f+f\sigma_8$, OHD+SNIa+$f\sigma_8$, OHD+SNIa+$\sigma_8+f$,  OHD+SNIa+$\sigma_8+f+f\sigma_8 $ datasets. Also, $\Delta$BIC values are less than 6, this indicates that the $f(Q)$ gravity model has ranges positively against $\Lambda$CDM, i.e., not completely ruled out. }

\section{Conclusions}
\label{gravity2} 
In the context of the nonmetric modified theory of gravity, we have performed a thorough statistical analysis resorting to both cosmological expansion history and large-scale structure data. Throughout the investigation, we considered a generic class of $f(Q)$ power-law models whose gravitational Lagrangian is of the form $f(Q) = Q+\alpha\, Q^n$. 

First, in order to study the cosmic history, in Section \ref{gravity} MCMC simulations were employed to estimate the best-fit outcomes for the parameters $\Omega_m$, $H_0$, and $n$. The results were presented in Table \ref{Tableone}, drawing upon data from  OHD, SNIa, and OHD+SNIa measurements. Therein OHD data provide their best fit for exponent $n>0$, whereas both SNIa and SNIa+OHD provide a similar negative value of $n$ as their best fit.
%
Then, an extensive statistical analysis, encompassing ${\mathcal{L}(\hat{\theta}|data)}$, $\chi^2$, $\chi^2$-red(uced), $\rm{AIC}$, $\Delta \rm{AIC}$, $\rm{BIC}$, and $\Delta \rm{BIC}$, was carried out (see Table \ref{stastical1}). 
It turned out that for the OHD, SNIa, and OHD+SNIa datasets, the $\Delta \rm{AIC}$ estimator provided values of {$1.46$, $1.74$, and $1.42$}, respectively, indicating robust support for the proposed $f(Q)$ paradigm. However, the  $\Delta \rm{BIC}$ values corresponded to {$3.21$, $7.20$, and $5.88$}, respectively, suggesting comparatively weaker support for the power-law $f(Q)$ gravity model. This apparent contradiction prompted us to assess the competitiveness of the chosen $f(Q)$ models resorting to further statistical investigation. The way to follow consisted of utilizing both cosmic expansion measurements and large-scale structure data.

Next, in Section \ref{gravity1} we provided a detailed analysis of the scalar perturbations equations in $f(Q)$ gravity resorting to the $1+3$ covariant formalism.  As such, the spatial gradients of the evolution equations were presented, and scalar and harmonic decomposition techniques were applied to extract the matter density fluctuations. After having derived the evolution equations, both for the full system without any approximation and their counterparts within the quasi-static approximation, we considered a matter-dominated universe to present the numerical results of the density fluctuations corresponding to both the whole system, i.e., with no simplifying approximation, governed by Eq. \eqref{second2} - shown in the left panel of Fig. \ref{fig:q1} - and the quasi-static approximation by means of Eq. \eqref{rr}  - shown in the right panel of Fig. \ref{fig:q1} - for statistically preferred values of the exponent $n$.  Additionally, our analysis allowed us to both determine and quantify the discrepancy of theoretical predictions issued from the $\Lambda$CDM and $f(Q)$ power-law model when the density contrast is the variable of interest.
To do so, 
we introduced the dimensionless parameter $\zeta$ as provided in Eq. \eqref{varaince}. The numerical results of $\zeta(z)$ were also presented in the inner left and right panels of Fig. \ref{fig:q1} to see the corresponding amplitude deviations of $\delta(z)$ between  $\Lambda$CDM and $f(Q)$ power-law model by considering either the solution of the full system or that within the quasi-static approximation. We were also capable of comparing the density fluctuations predictions when obtained from either the quasi-static approximation or the full system, as represented in Fig. \ref{fig:qqxx}. Such a figure served us to claim that in the context of $f(Q)$ gravity power-law models, the quasi-static approximation
provides a faithful representation of the full resolution, with consistency levels of {  94.03\% at $n = 0.215$, 98.7\% at $n = -0.085$, and 96.61\% at $n = -0.275$.}

The growth of structures in $f(Q)$ gravity was then compared with either redshift-space distortion data $f\sigma_8$ or some recent separate measurements of the growth rate $f(z)$ and the amplitude of matter fluctuations $\sigma_8(z)$ in Section \ref{sigma}. Therein, we found the best-fit values of cosmological parameters $\Omega_m$, $\sigma_8$ and $n$ using $\sigma_8+f$, $\sigma_8+f$ and $\sigma_8+f+f\sigma_8$ data, see Fig. \ref{fig:enter-label}.
Moreover, we also performed analyses resorting to OHD + SNIa + $f\sigma_8$, OHD + SNIa + $\sigma_8+f$ and OHD + SNIa + $\sigma_8+f\,+\,f\sigma_8$  data, see Fig. \ref{ohdsnfsigma81} and Table \ref{bestfitparameters}. 
Thus, we conclude that for all the analyses performed, except for $f\sigma_8$ and $f+\sigma_8+f\sigma_8$, the best-fit exponent $n$ is negative.
Subsequently, we performed a statistical analysis to demonstrate the goodness of the fits of the $f(Q)$ gravity models.
Our statistical analysis showed that for all datasets, except for $f\sigma_8$ and $\sigma_8+f$, $\Delta \rm{AIC}$ values 
indicated observational support. On the other hand, when the \rm{BIC} criterion is the one to be applied, it happened that $ 2\leq \Delta\rm{BIC} \leq 6$, for all datasets except $\sigma_8 + f$, i.e., the $f(Q)$ gravity is positive against to $\Lambda$CDM model. All of these results are summarized in Table \ref{Tab:values}. 
These results show that, at the present stage, using only three $f$ and three $\sigma_8$ data may affect the goodness of $f(Q)$ models against the $\Lambda$CDM, the latter taken as the reference model. This happens despite the fact that those $\sigma_8+f$ data noticeably reduce the contour plots areas - and therefore the allowed parameter values within 1$\sigma$ and 2$\sigma$ confidence levels. See Figs. \ref{fig:enter-label} and \ref{fig:matter-universe}. 
The use of further 
$\{f, \sigma_8\}$ data once available, as well as growth rate $f$ data - obtained independently from $\sigma_8$ through the analyses of a diversity of cosmological tracers, including luminous red galaxies, blue galaxies, voids, and quasars - may shed further light about the ability of such separate observations $\{f, \sigma_8\}$ to dismiss - or deepen - the aforementioned tension or consolidate the Concordance $\Lambda$CDM model. Anyhow, more of these $\{f, \sigma_8\}$ data separated will greatly enhance the power of growth of structures to probe departures from the usual Einsteinian gravity. Work in this direction is in progress.

\section*{Acknowledgments }

The authors would like to thank Javier de Cruz P\'erez, Renier Hough, and Antonio L. Maroto for insightful comments during the preparation of the manuscript. Endalkachew Tsegaye was initially involved in earlier aspects of the work, and his contribution is duly acknowledged.
AA acknowledges that this work was part of the research programme “New Insights into Astrophysics and Cosmology with Theoretical Models Confronting Observational Data” of the National Institute for
Theoretical and Computational Sciences of South Africa. 
AdlCD acknowledges support from BG20/00236 action (MCINU, Spain), NRF Grant CSUR23042798041, CSIC Grant COOPB23096, Project SA097P24 funded by Junta de Castilla y Le\'on (Spain) and Grant PID2021-122938NB-I00 funded by MCIN/AEI/10.13039/501100011033 and by {\it ERDF A way of making Europe}.

\section*{Data Availability}
There are no associated data with this article. No new data were generated or analysed in support of this research.
%
%
%



\bibliographystyle{mnras}
\bibliography{references} 



\clearpage
\appendix
\section{Growth structure data}\label{Appendix}

\begin{table}
{

\centering
\begin{tabular}{|c c c c c|} 
 \hline
 Dataset & $z$ &  $f$  &$\sigma_8$&Ref.\\ 
 SDSS & 0.10 &  0.464 $\pm$ 0.040& 0.769 $\pm$ 0.105& \cite{shi2019mapping}\\
 Vipers PDR-2&0.60&0.93 $\pm$ 0.22& 0.52 $\pm$ 0.06&\cite{De11,shambel}\\
 Vipers PDR-2&0.86&0.99 $\pm$ 0.19&0.48 $\pm$ 0.04&\cite{De11,shambel}\\
 
 \hline
\end{tabular}
\caption{Three data points for growth rate ($f$), and three data points for the normalization of the matter power spectrum at the scale of $8h^{-1}{\rm Mpc}$ ($\sigma_8$).}
\label{datasetsofgrowth1}
\begin{tabular}{|c c c  c|} 
 \hline
 Dataset & $z$ & $f\sigma_8$  &Ref.\\ 
 \hline
 2MTF & 0.001 & 0.505 $\pm$ 0.085 & \cite{Howlett}\\ 
 6dFGS+SNIa & 0.02& 0.428 $\pm$ 0.0465 &\cite{HutererD} \\
 IRAS+SNIa & 0.02 & 0.398 $\pm$ 0.065 &  \cite{HudsonMJ,TurnbullSJ}\\
 2MASS & 0.02 & 0.314 $\pm$ 0.048 & \cite{HudsonMJ,Davis}\\
 SDSS & 0.10 & 0.376 $\pm$ 0.038 &  \cite{shi2019mapping}\\
 SDSS-MGS &0.15&0.490$\pm$ 0.145&\cite{HowlettC}\\
 2dFGRS &0.17&0.510 $\pm$ 0.060&\cite{SongYS}\\
 GAMA &0.18&0.360 $\pm$ 0.090&\cite{BlakeC1}\\
 GAMA&0.38&0.440 $\pm$ 0.060& \cite{BlakeC1}\\
 SDSS-LRG-200&0.25&0.3512 $\pm$ 0.0583&\cite{SamushiaL}\\
 SDSS-LRG-200&0.37&0.4602 $\pm$ 0.0378&\cite{SamushiaL}\\
 BOSS DR12&0.31&0.469 $\pm$ 0.098&\cite{YWang}\\
 BOSS DR12&0.36&0.474 $\pm$ 0.097&\cite{YWang}\\
 BOSS DR12&0.40&0.473 $\pm$ 0.086&\cite{YWang}\\
 BOSS DR12&0.44&0.481 $\pm$ 0.076&\cite{YWang}\\
 BOSS DR12&0.48&0.482 $\pm$ 0.067&\cite{YWang}\\
 BOSS DR12&0.52&0.488 $\pm$ 0.065&\cite{YWang}\\
 BOSS DR12&0.56&0.482 $\pm$ 0.067&\cite{YWang}\\
 BOSS DR12&0.59&0.481 $\pm$ 0.066&\cite{YWang}\\
 BOSS DR12&0.64&0.486 $\pm$ 0.070&\cite{YWang}\\
 WiggleZ&0.44&0.413 $\pm$ 0.080&\cite{BlakeC}\\
 WiggleZ&0.60&0.390 $\pm$ 0.063& \cite{BlakeC}\\
 WiggleZ&0.73&0.437 $\pm$ 0.072&\cite{BlakeC}\\
 Vipers PDR-2&0.60&0.550 $\pm$ 0.120&\cite{De11,shambel}\\
 Vipers PDR-2&0.86&0.400 $\pm$ 0.110&\cite{De11,shambel}\\
 FastSound&1.40&0.482 $\pm$ 0.116&\cite{okumura2016subaru}\\
 SDSS-IV&0.978&0.379 $\pm$ 0.176&\cite{wang2018clustering} \\
 SDSS-IV&1.23&0.385 $\pm$ 0.099&\cite{wang2018clustering}\\
 SDSS-IV&1.526&0.342 $\pm$ 0.070&\cite{wang2018clustering}\\
 
 SDSS-IV&1.944&0.364 $\pm$ 0.106&\cite{wang2018clustering}\\
 \hline
\end{tabular}
\caption{This study incorporates a collection of structure data, consisting of 30 data points for redshift space distortion ($f\sigma_8$), 3 data points for growth rate ($f$), and 3 data points for the normalization of the matter power spectrum at the scale of $8h^{-1}{\rm Mpc}$ ($\sigma_8$).
}
 \label{datasetsofgrowth}
 }
\end{table}

\end{document}